\documentclass[12pt,a4paper]{article}
\usepackage{graphics,graphicx}
\usepackage{amssymb,epsfig,amsmath,euscript,array}
\usepackage{cite}

\makeatletter
\@addtoreset{equation}{section}
\makeatother
\renewcommand{\theequation}{\thesection.\arabic{equation}}

\newcommand{\startappendix}{
\setcounter{section}{0}
\renewcommand{\thesection}{\Alph{section}}
\renewcommand{\theequation}{\Alph{section}.\arabic{equation}}}

\newcommand{\Appendix}[1]{
\refstepcounter{section}
\begin{flushleft}
{\Large\bf Appendix \thesection: #1}
\end{flushleft}}

\newcounter{multieqs}

\newcommand{\be}{\begin{equation}}
\newcommand{\ee}{\end{equation}}

\newcommand{\bm}[1]{\mbox{\boldmath $#1$}}

\newcommand{\kslash}{k \!\!\! / }

\newcommand{\lslash}{l \!\! / }
\newcommand{\Pslash}{P \!\!\!\! / }

\newcommand{\islash}{i \!\!\! / }
\newcommand{\jslash}{j \!\!\! / }
\newcommand{\aslash}{a \!\!\! / }
\newcommand{\bslash}{{b \hspace{-6pt} \slash} }

\newcommand{\onslash}{1 \!\!\! / }
\newcommand{\twslash}{2 \!\!\!/ }
\newcommand{\thslash}{3 \!\!\!/ }
\newcommand{\foslash}{4 \!\!\! / }
\newcommand{\fislash}{5 \!\!\! / }

\newcommand{\mslash}{m \!\!\! / }

\def\bd{\begin{document}}
\def\ed{\end{document}}
\def\nn{\nonumber}
\def\bea{\begin{eqnarray}}
\def\eea{\end{eqnarray}}
\def\eps{\epsilon}

\def\ab{(ijab)}
\def\ba{(ijba)}
\def\ijab{{\tr}_{+}(\islash\, \jslash\, \aslash \, \bslash)}
\def\ijba{{\tr}_{+}(\islash\, \jslash\, \bslash \, \aslash)}
\def\ijaP{{\tr}_{+}(\islash\, \jslash\, \aslash \, \Pslash)}
\def\ijPLa{{\tr}_{+}(\islash\, \jslash\, \Pslash_L \, \aslash)}
\def\ijaPL{{\tr}_{+}(\islash\, \jslash\, \aslash \, \Pslash_L)}
\def\ijPLza{{\tr}_{+}(\islash\, \jslash\, \Pslash_{L;z} \, \aslash)}
\def\ijaPLz{{\tr}_{+}(\islash\, \jslash\, \aslash \, \Pslash_{L;z})}
\def\ijPa{{\tr}_{+}(\islash\, \jslash\, \Pslash \, \aslash)}
\def\iaPb{{\tr}_{+}(\islash\, \aslash\, \Pslash \, \bslash)}
\def\ibPa{{\tr}_{+}(\islash\, \bslash\, \Pslash \, \aslash)}
\def\ijPmu{{\tr}_{+}(\islash\, \jslash\, \Pslash \, \mu)}
\def\ibmuP{{\tr}_{+}(\islash\, \bslash\, \mu \, \Pslash)}
\def\ibmua{{\tr}_{+}(\islash\, \bslash\, \mu \, \aslash)}
\def\iamub{{\tr}_{+}(\islash\, \aslash\, \mu \, \bslash)}
\def\jaPb{{\tr}_{+}(\jslash\, \aslash\, \Pslash \, \bslash)}
\def\ijmuP{{\tr}_{+}(\islash\, \jslash\, \mu \, \Pslash)}
\def\ijmum{{\tr}_{+}(\islash\, \jslash\, \mu \, \mslash)}
\def\ijmmu{{\tr}_{+}(\islash\, \jslash\, \mslash \, \mu)}
\def\ijmP{{\tr}_{+}(\islash\, \jslash\, \mslash \, \Pslash)}
\def\iabP{{\tr}_{+}(\islash\, \aslash\, \bslash \, \Pslash)}
\def\ijbP{{\tr}_{+}(\islash\, \jslash\, \bslash \, \Pslash)}
\def\jbPa{{\tr}_{+}(\jslash\, \bslash\, \Pslash \, \aslash)}
\def\ijPb{{\tr}_{+}(\islash\, \jslash\, \Pslash \, \bslash)}
\def\jbmua{{\tr}_{+}(\jslash\, \bslash\, \mu \, \aslash)}

\def\loablt{ {\tr}_{+}(\lslash_1\, \aslash \, \bslash\, \lslash_2)}

\def\ijlolt{{\tr}_{+}(\islash\, \jslash\, \lslash_1 \, \lslash_2)}
\def\ijltlo{{\tr}_{+}(\islash\, \jslash\, \lslash_2 \, \lslash_1)}
\def\ibloa{{\tr}_{+}(\islash\, \bslash\, \lslash_1 \, \aslash)}
\def\jaltb{{\tr}_{+}(\jslash\, \aslash\, \lslash_2 \, \bslash)}
\def\ialtb{{\tr}_{+}(\islash\, \aslash\, \lslash_2 \, \bslash)}
\def\bltloa{{\tr}_{+}(\bslash\, \lslash_2\, \lslash_1 \, \aslash)}
\def\jbloa{{\tr}_{+}(\jslash\, \bslash\, \lslash_1 \, \aslash)}
\def\ibPb{{\tr}_{+}(\islash\, \bslash\, \Pslash \, \bslash)}
\def\ijltb{{\tr}_{+}(\islash\, \jslash\, \lslash_2 \, \bslash)}

\def\ijloa{{\tr}_{+}(\islash\, \jslash\,  \lslash_1 \, \aslash)}
\def\ijblt{{\tr}_{+}(\islash\, \jslash\,  \bslash \, \lslash_2)}

\def\jakb{{\tr}_{+}(\jslash\, \aslash\, \kslash \, \bslash)}
\def\iakb{{\tr}_{+}(\islash\, \aslash\, \kslash \, \bslash)}

\def\tofo{{\tr}_{+}(\onslash\, \thslash\, \twslash \, \foslash)}
\def\foto{{\tr}_{+}(\onslash\, \thslash\, \foslash \, \twslash)}
\def\tofi{{\tr}_{+}(\onslash\, \thslash\, \twslash \, \fislash)}
\def\fito{{\tr}_{+}(\onslash\, \thslash\, \fislash \, \twslash)}

\def\lrangle#1#2{\langle #1\,#2\rangle}

\def\Li{{$\rm Li}_2$}

\let\bm=\bibitem
\let\la=\label

\def\npb#1#2#3{Nucl. Phys. {\bf{B#1}} #3 (#2)}
\def\plb#1#2#3{Phys. Lett. {\bf{#1B}} #3 (#2)}
\def\prl#1#2#3{Phys. Rev. Lett. {\bf{#1}} #3 (#2)}
\def\prd#1#2#3{Phys. Rev. {D \bf{#1}} #3 (#2)}
\def\cmp#1#2#3{Comm. Math. Phys. {\bf{#1}} #3 (#2)}
\def\cqg#1#2#3{Class. Quantum Grav. {\bf{#1}} #3 (#2)}
\def\nppsa#1#2#3{Nucl. Phys. B (Proc. Suppl.) {\bf{#1A}}#3 (#2)}
\def\ap#1#2#3{Ann. of Phys. {\bf{#1}} #3 (#2)}
\def\ijmp#1#2#3{Int. J. Mod. Phys. {\bf{A#1}} #3 (#2)}
\def\rmp#1#2#3{Rev. Mod. Phys. {\bf{#1}} #3 (#2)}
\def\mpla#1#2#3{Mod. Phys. Lett. {\bf A#1} #3 (#2)}
\def\jhep#1#2#3{J. High Energy Phys. {\bf #1} #3 (#2)}
\def\atmp#1#2#3{Adv. Theor. Math. Phys. {\bf #1} #3 (#2)}
\newcommand{\EQ}[1]{\begin{equation} #1 \end{equation}}
\newcommand{\AL}[1]{\begin{subequations}\begin{align} #1 \end{align}\end{subequations}}
\newcommand{\SP}[1]{\begin{equation}\begin{split} #1 \end{split}\end{equation}}
\newcommand{\ALAT}[2]{\begin{subequations}\begin{alignat}{#1} #2 \end{alignat}
                        \end{subequations}}
\def\beqa{\begin{eqnarray}}
\def\eeqa{\end{eqnarray}}
\def\beq{\begin{equation}}
\def\eeq{\end{equation}}
\def\sst{\scriptscriptstyle}
\def\thetabar{\bar\theta}
\def\Tr{{\rm Tr}}
\def\one{\mbox{1 \kern-.59em {\rm l}}}
 \def\Nh{\hat{N}}

\def\a{\alpha}      \def\da{{\dot\alpha}}
\def\b{\beta}       \def\db{{\dot\beta}}
\def\g{\gamma}  \def\G{\Gamma}  \def\cdt{\dot\gamma}
\def\d{\delta}  \def\D{\Delta}  \def\ddt{\dot\delta}
\def\e{\epsilon}        \def\vare{\varepsilon}
\def\f{\phi}    \def\F{\Phi}    \def\vvf{\f}
\def\h{\eta}
\def\k{\kappa}
\def\l{\lambda} \def\L{\Lambda}
\def\m{\mu} \def\n{\nu}
\def\o{\omega}
\def\p{\pi} \def\P{\Pi}
\def\r{\rho}
\def\s{\sigma}  \def\S{\Sigma}
\def\t{\tau}
\def\th{\theta} \def\Th{\Theta} \def\vth{\vartheta}
\def\X{\Xeta}
\def\z{\zeta}

\def\cA{{\cal A}} \def\cB{{\cal B}} \def\cC{{\cal C}}
\def\cD{{\cal D}} \def\cE{{\cal E}} \def\cF{{\cal F}}
\def\cG{{\cal G}} \def\cH{{\cal H}} \def\cI{{\cal I}}
\def\cJ{{\cal J}} \def\cK{{\cal K}} \def\cL{{\cal L}}
\def\cM{{\cal M}} \def\cN{{\cal N}} \def\cO{{\cal O}}
\def\cP{{\cal P}} \def\cQ{{\cal Q}} \def\cR{{\cal R}}
\def\cS{{\cal S}} \def\cT{{\cal T}} \def\cU{{\cal U}}
\def\cV{{\cal V}} \def\cW{{\cal W}} \def\cX{{\cal X}}
\def\cY{{\cal Y}} \def\cZ{{\cal Z}}

\def\ua{\underline{\alpha}}
\def\ub{\underline{\phantom{\alpha}}\!\!\!\beta}
\def\uc{\underline{\phantom{\alpha}}\!\!\!\gamma}
\def\um{\underline{\mu}}
\def\ud{\underline\delta}
\def\ue{\underline\epsilon}
\def\una{\underline a}\def\unA{\underline A}
\def\unb{\underline b}\def\unB{\underline B}
\def\unc{\underline c}\def\unC{\underline C}
\def\und{\underline d}\def\unD{\underline D}
\def\une{\underline e}\def\unE{\underline E}
\def\unf{\underline{\phantom{e}}\!\!\!\! f}\def\unF{\underline F}
\def\unm{\underline m}\def\unM{\underline M}
\def\unn{\underline n}\def\unN{\underline N}
\def\unp{\underline{\phantom{a}}\!\!\! p}\def\unP{\underline P}
\def\unq{\underline{\phantom{a}}\!\!\! q}
\def\unQ{\underline{\phantom{A}}\!\!\!\! Q}
\def\unH{\underline{H}}

\def\As {{A \hspace{-6.4pt} \slash}\;}
\def\bs {{b \hspace{-6.4pt} \slash}\;}
\def\Ds {{D \hspace{-6.4pt} \slash}\;}
\def\ds {{\del \hspace{-6.4pt} \slash}\;}
\def\ss {{\s \hspace{-6.4pt} \slash}\;}
\def\ks {{ k \hspace{-6.4pt} \slash}\;}
\def\ps {{p \hspace{-6.4pt} \slash}\;}
\def\pas {{{p_1} \hspace{-6.4pt} \slash}\;}
\def\pbs {{{p_2} \hspace{-6.4pt} \slash}\;}
\def\Ps {{P \hspace{-6.4pt} \slash}\;}
\def\Qs {{Q \hspace{-6.4pt} \slash}\;}

\def\Fh{\hat{F}}
\def\Vh{\hat{V}}
\def\Xh{\hat{X}}
\def\ah{\hat{a}}
\def\xh{\hat{x}}
\def\yh{\hat{y}}
\def\ph{\hat{p}}
\def\xih{\hat{\xi}}
\def\psit{\tilde{\psi}}
\def\Psit{\tilde{\Psi}}
\def\tht{\tilde{\th}}
\def\lt{\tilde{\lambda}}
\def\llt{\tilde{l}}
\def\At{\tilde{A}}
\def\Qt{\tilde{Q}}
\def\Rt{\tilde{R}}
\def\Nt{\tilde{N}}

\def\at{\tilde{a}}
\def\st{\tilde{s}}
\def\ft{\tilde{f}}
\def\pt{\tilde{p}}
\def\qt{\tilde{q}}
\def\vt{\tilde{v}}
\def\nt{\tilde{n}}

\def\delb{\bar{\partial}}
\def\bz{\bar{z}}
\def\bD{\bar{D}}
\def\bB{\bar{B}}

\def\bk{{\bf k}}
\def\bl{{\bf l}}
\def\bp{{\bf p}}
\def\bq{{\bf q}}
\def\br{{\bf r}}
\def\bx{{\bf x}}
\def\by{{\bf y}}
\def\bR{{\bf R}}
\def\bV{{\bf V}}

\def\d{\delta}\def\D{\Delta}\def\ddt{\dot\delta}
\def\pa{\partial} \def\del{\partial}
\def\xx{\times}
\def\uno{\mbox{1 \kern-.59em {\rm l}}}
\def\trp{^{\top}}
\def\inv{^{-1}}
\def\dag{{^{\dagger}}}
\def\pr{^{\prime}}
\def\lan{\langle}
\def\ran{\rangle}
\def\rar{\rightarrow}
\def\lar{\leftarrow}
\def\lrar{\leftrightarrow}
\newcommand{\0}{\,\!}      
\def\one{1\!\!1\,\,}
\def\im{\imath}
\def\jm{\jmath}
\newcommand{\tr}{\mbox{tr}}
\newcommand{\slsh}[1]{/ \!\!\!\! #1}
\def\vac{|0\rangle}
\def\lvac{\langle 0|}
\def\hlf{\frac{1}{2}}
\def\ove#1{\frac{1}{#1}}
\def\Box{\square}
\def\ZZ{\mathbb{Z}}
\def\CC#1{({\bf #1})}
\def\bcomment#1{}
\def\bfhat#1{{\bf \hat{#1}}}
\def\VEV#1{\left\langle #1\right\rangle}
\newcommand{\ex}[1]{{\rm e}^{#1}} \def\ii{{\rm i}}
\def\rr{{\rm r}} \def\rs{{\rm s}}\def\rv{{\rm v}}
\def\ri{{\rm i}}\def\rj{{\rm j}}
\newcommand{\lrbrk}[1]{\left(#1\right)}
\newcommand{\sfrac}[2]{{\textstyle\frac{#1}{#2}}}

\def\Li{{\rm Li}_2}

\font\mybb=msbm10 at 12pt
\def\bb#1{\hbox{\mybb#1}}

\font\myBB=msbm10 at 18pt
\def\BB#1{\hbox{\myBB#1}}

\begin{document}

\begin{flushright}
hep-th/0506068 \\
QMUL-PH-05-09
\end{flushright}

\vspace{20pt}

\begin{center}

{\Large \bf Loop Amplitudes in Pure Yang-Mills }
\\
\vspace{0.3cm} {\Large \bf  from Generalised Unitarity  } \vspace{12pt}
\vspace{33pt}

{\bf {\mbox{Andreas  Brandhuber, Simon McNamara, Bill
Spence and Gabriele  Travaglini}}}%
\footnote{
{\sffamily \{\tt a.brandhuber, s.mcnamara, w.j.spence,
g.travaglini\}@qmul.ac.uk }}

{\em Department of Physics\\
Queen Mary, University of
London\\
Mile End Road, London, E1 4NS\\
United Kingdom}
\vspace{40pt}

{\bf Abstract}

\end{center}


\noindent
We show how generalised unitarity cuts in $D\!=\! 4 - 2 \e$
dimensions can be used to
calculate efficiently complete
one-loop scattering amplitudes in
non-supersymmetric Yang-Mills theory. This
approach naturally generates the rational terms in the amplitudes,
as well as the cut-constructible parts.
We test the validity of our
method by re-deriving the one-loop
$+$$+$$+$$+$, $-$$+$$+$$+$, $-$$-$$+$$+$, $-$$+$$-$$+$
and $+$$+$$+$$+$$+$
gluon scattering amplitudes
using generalised quadruple cuts
and triple cuts in $D$ dimensions.

\vspace{0.5cm}

\setcounter{page}{0}
\thispagestyle{empty}
\newpage


                     \section{Introduction}


\setcounter{footnote}{0}

Over the past year, major progress in the calculation
of scattering amplitudes in perturbative Yang-Mills theory
has been made. This  was triggered by
Witten's discovery that tree-level amplitudes in Yang-Mills
can equivalently be derived via a string theory calculation,
where the string theory in question is the topological
B model with target space a supersymmetric version of
Penrose's twistor space \cite{witten}.
Witten also observed that tree-level scattering amplitudes,
when Fourier transformed to twistor space, have
an interesting geometrical structure, namely they have
support on algebraic curves; for the simple case of the
maximally helicity violating (MHV) amplitude,
described by the Parke-Taylor formula,
the curve is just a line (for real twistor space).
This remarkable observation gives an explanation for
the unexpected and previously rather mysterious simplicity
of tree-level scattering amplitudes in Yang-Mills
such as the Parke-Taylor formula,
which is not at all apparent in a calculation
performed using standard Feynman rules.

On a different line of development, the simplicity of
tree-level scattering amplitudes was linked to the existence
of novel recursion relations discovered by
Britto, Cachazo and Feng (BCF) \cite{bcf}, and subsequently proved
by the same authors and Witten  (BCFW) \cite{bcfw}.
The elegant proof of \cite{bcfw} is based on very general
properties of amplitudes,  such as analyticity
\cite{Landau:1959fi,Cutkosky:1960sp,bible}
and factorisation on multiparticle poles, and hence
gave rise to the hope that recursion relations may
arise in very different contexts.
Indeed,  novel recursion relations were also found
in general relativity \cite{bbst3,cs}, scalar theory \cite{bbst3},
for the finite rational amplitudes at one-loop in Yang-Mills
and massless QCD
\cite{bdk-rational1,bdk-rational2}, and for tree amplitudes
involving massive scalars and gluons in Yang-Mills \cite{snvp}.

The simplicity of tree-level amplitudes
in Yang-Mills was exploited by Bern, Dixon, Dunbar and Kosower
(BDDK) in order to build one-loop scattering amplitudes
\cite{bdk1,bdk2}.
By applying unitarity at the level of amplitudes, rather than
Feynman diagrams, these authors were able to construct
many one-loop amplitudes in supersymmetric theories,
such as the infinite sequence of MHV amplitudes in
$\cN\! = \! 4$ and in $\cN\! = \! 1$ super Yang-Mills
(SYM).
The unitarity method of BDDK by-passes the use
of Feynman diagrams
and its related complications,
and generates  results of an unexpectedly simple form;
for instance, the one-loop MHV amplitude in $\cN\! = \! 4$ SYM
is simply given by the tree-level expression
multiplied by a sum of ``two-mass easy'' box functions,
all with coefficient one.
As a side remark, we would like  to mention that
higher-loop amplitudes in $\cN \! = \! 4$ SYM also
display  intriguing regularities \cite{babis1,babis2,bds}.

The geometrical structure in twistor space of the amplitudes
was also the root of a further important development.
In \cite{csw}, Cachazo, Svr\v{c}ek  and Witten
(CSW) proposed a novel perturbative expansion for
on-shell amplitudes in Yang-Mills,
where the MHV amplitudes are lifted to vertices,
joined by simple scalar
propagators in order to  form amplitudes with
an increasing number of negative helicities.
Applications at tree level
confirmed the validity of the method and led to
the derivation of various new amplitudes in gauge theory
\cite{csw,gv,zhu,wu-zhu1,wu-zhu2,gnv,lnv,Bern:2004ba}.

In \cite{csw}, a heuristic derivation of the
CSW method was given from the twistor string theory.
Rather unfortunately, the latter only appears to describe
the scattering amplitudes of Yang-Mills at tree level
\cite{bw}, as at one loop states of conformal supergravity
enter the game, and cannot
be decoupled in any known limit. The duality
between gauge theory and twistor string theory
is thus spoiled by quantum corrections.
Surprisingly,
it was found by three of the present authors that the MHV method
at one-loop level does succeed in correctly reproducing
the scattering amplitudes of the  gauge theory
\cite{bst}. Furthermore, the twistor space picture of
one-loop amplitudes is now in complete agreement with that
emerging from the MHV methods, which suggests that the amplitudes at
one loop have localisation properties
on unions of lines in twistor space;
an initial puzzle \cite{csw2} was indeed clarified and explained
in terms of a certain ``holomorphic anomaly'', introduced in
\cite{csw3}, and further analysed in
\cite{lits,Cachazo:2004dr,Britto:2004nj,Bern:2004ky,Bidder:2004tx}.
A proof of the MHV method at tree level was finally given
in \cite{bcfw}; at loop level, however, it remains a
(well-supported) conjecture.

The initial successful application of the MHV method
to $\cN\!=\!4$ SYM \cite{bst} was followed by
calculations of MHV amplitudes in
$\cN\!=\!1$ SYM \cite{qr,bbst},
and in pure Yang-Mills \cite{bbst2},
where the four-dimensional cut-constructible part
of the infinite sequence of
MHV amplitudes was derived.
However, amplitudes in non-supersymmetric Yang-Mills theory
also have  rational terms which escape analyses
based on MHV diagrams at one loop
\cite{bst,bbst2} or four-dimensional unitarity
\cite{bdk1,bdk2}.

Amplitudes in supersymmetric theories are
of course  special.
They do contain rational terms, but these are uniquely linked
to terms which have cuts in four dimensions.
In other words, these amplitudes can  be reconstructed
uniquely from their cuts in four-dimensions \cite{bdk1,bdk2} --
a remarkable result.
These cuts are of course four-dimensional tree-level
amplitudes, whose simplicity is instrumental
in allowing the derivation of analytic, closed-form
expressions for the one-loop amplitudes.
In non-supersymmetric theories, amplitudes can still
be reconstructed from their cuts, but
on the condition of working in
$4-2 \e$ dimensions, with $\e \neq 0$
\cite{vanNeerven:1985xr,Bern:1995db,Bern:1996ja}.
This is a powerful statement, but it also implies
the rather unpleasant fact that one should in principle
work with
tree-level amplitudes involving gluons continued to
$4-2 \e$ dimensions, which are not simple.

An important simplification is offered by the well-known
supersymmetric decomposition of one-loop amplitudes of gluons
in pure Yang-Mills.
Given a one-loop amplitude $\cA_{\rm g}$
with gluons running the loop,
one can re-cast it as
\beq
\label{susydec}
\cA_{\rm g} \ = \ (\cA_{\rm g} \, + \, 4 \cA_{\rm f} \, + \,
3 \cA_{\rm s}) \  - \
4( \cA_{\rm f}  + \cA_{\rm s}) \ + \  \cA_{\rm s}
\ .
\eeq
Here $\cA_{\rm f}$ ($\cA_{\rm s}$) is the amplitude with the same
external particles as   $\cA_{\rm g}$ but with a Weyl fermion
(complex scalar) in the adjoint of the gauge group running
in the loop.
This decomposition is useful because
the first two terms on the right hand side of
\eqref{susydec} are  contributions
coming from an $\cN \! = \! 4$ multiplet and
(minus four times) a chiral $\cN \! = \! 1$ multiplet,
respectively; therefore, these terms are
four-dimensional cut-constructible,
which simplifies their calculation enormously.
The last term in \eqref{susydec}, $\cA_{\rm s}$,
is the contribution coming from a scalar running in the loop.
The key point here is that the calculation
of this term is much easier than that
of the original amplitude $\cA_{\rm g}$.
It is this last contribution which is the focus of this paper.

The root of the simplification lies in the fact that
a massless scalar in $4-2\e$ dimensions can equivalently
be described  as a massive scalar in four dimensions
\cite{Bern:1995db,Bern:1996ja}.
Indeed, if $L$ is the $(4-2\e)$-dimensional momentum of the
massless scalar ($L^2 \!=\! 0$), decomposed into a
four-dimensional component $l_{(4)}$ and a $ -2\e$-dimensional
component $l_{(-2 \e)}$,
$L := l_{(4)} + l_{(-2\e)}$, one has
$L^2 := l_{(4)}^2 +  l_{(-2\e)}^2 = l_{(4)}^2 - \mu^2$,
where $l_{(-2\e)}^2 := - \mu^2$ and the four-dimensional
and $-2\e$-dimensional subspaces are taken to be orthogonal.
The tree-level amplitudes entering the $(4-2\e)$-dimensional
cuts of a one-loop amplitude with a scalar in the loop
are therefore those involving a pair of
massive scalars and gluons.
Crucially, these  amplitudes have a rather  simple form.
Some of these amplitudes appear in \cite{Bern:1995db,Bern:1996ja};
furthermore, a recent paper \cite{snvp} describes
how to efficiently derive such amplitudes using
a recursion relation similar to that of BCFW.

Using two-particle cuts in $4-2\e$ dimensions, together with
the supersymmetric decomposition mentioned above,
various amplitudes in pure Yang-Mills
were derived in recent years, starting with the
pioneering works
\cite{Bern:1995db,Bern:1996ja}.
In this paper we show that
this  analysis can be performed with the help of
an additional tool: generalised  $(4-2\e)$-dimensional unitarity.

Generalised four-dimensional unitarity
\cite{Cutkosky:1960sp,bible,a,b,c}
was very efficiently applied in \cite{bcf-gen}
to the calculation of one-loop amplitudes
in $\cN \! = \! 4$ SYM. Amplitudes in this theory
can be written as a sum of box functions, multiplied by rational
coefficients. To each box function is uniquely associated
a (generalised) quadruple cut, so that, schematically,
each coefficient of a box function is expressed as
a particular quadruple cut of the one-loop amplitude,
which is nothing but a product of four tree-level amplitudes.
Generalised cuts require the amplitudes to be continued to
complexified Minkowski space, which in turn has
the consequence that three-point amplitudes no longer vanish,
and enter the cut-amplitude
in an important way \cite{bcf-gen}.
\footnote{This circumstance extends to the $(4-2\e)$-dimensional
three-point scattering amplitudes which will be considered
in this paper.}
The calculation of one-loop amplitudes in  $\cN \! = \! 4$ SYM
was in this way turned into an algebraic problem \cite{bcf-gen}.
Using generalised unitarity in four dimensions,
the infinite sequence of next-to-MHV  amplitudes
in $\cN \! = \! 4$ SYM was determined \cite{bdk-december};
generalised unitarity was also applied to $\cN \! = \! 1$ SYM,
in particular to the calculation of the next-to-MHV amplitude
with adjacent negative-helicity gluons \cite{BBDP}.
These amplitudes can be expressed solely in terms of
triangles, and were efficiently computed in
\cite{BBDP} using triple cuts.%
\footnote{A new calculation based on localisation in
spinor space was also introduced in
\cite{BBCF}.}

The main point of this paper is the observation that
generalised unitarity is actually
a useful concept also in $4-2\e$ dimensions; in turn
this means that generalised
$(4-2\e)$-dimensional unitarity is relevant for the calculation of
non-supersymmetric amplitudes at one loop.
In particular in this paper
we will be able to compute amplitudes in
non-supersymmetric Yang-Mills by using quadruple and triple
cuts in $4-2\e$ dimensions.
This is advantageous for at least three reasons.
First of all, working with multiple cuts
simplifies considerably the algebra, because
several on-shell conditions can be used
at the same time;
furthermore, for the case of quadruple cuts
the integration is actually completely frozen \cite{bcf-gen}
so that the coefficient of the relevant box functions
entering the amplitude can be calculated
without performing any integration at all.
Lastly, the tree-level sub-amplitudes which are sewn
together in order to
form the multiple cut of the amplitude are simpler
than those entering the two-particle cuts of the same
amplitude.
It seems clear that immediate further progress with this
approach will not require major new conceptual advances, and
that it will be directly applicable to more complicated
and currently unknown amplitudes.

We describe this method in some detail in Section 2,
and then move on to present various examples of its application.
Specifically, using generalised unitarity in $4-2\e$ dimensions
we will re-calculate the all-orders in $\e$ expressions of
all one-loop, four gluon scattering amplitudes
in non-supersymmetric Yang-Mills,
that is $+$$+$$+$$+$, $-$$+$$+$$+$, and the two MHV amplitudes
$-$$-$$+$$+$ and $-$$+$$-$$+$; and finally, the five-gluon
all-plus helicity amplitude $+$$+$$+$$+$$+$.
These amplitudes have already been computed
to all orders in $\e$ in \cite{Bern:1995db},
and we find in all cases complete agreement
with the results of that paper.
The examples we consider are complementary,
as they show that this method can be applied to
finite amplitudes without infrared divergences,
as well as to infrared divergent amplitudes containing both
rational and cut-constructible terms.
These calculations are described in Section 3 and Section 4.
In an Appendix we have collected some useful
definitions and formulae.


\section{Generalised Unitarity in $D=4\!-\!2 \eps$ Dimensions}


Conventional unitarity and generalised unitarity in {\it four dimensions}
have been shown to be extremely powerful tools
for calculating one-loop and higher-loop scattering amplitudes in
supersymmetric gauge theories and gravity.
At one-loop, conventional unitarity
amounts to reconstructing the full
amplitude from the knowledge of the
discontinuity or imaginary part of the amplitude.
In this process the
amplitude is cut into two tree-level, on-shell
amplitudes defined in four dimensions,
and the two propagators connecting
the two sub-amplitudes are replaced
by on-shell delta-functions
which reduce the loop integration to a phase
space integration. In principle this cutting
technique is only sensitive to terms in the amplitude that have
discontinuities, like logarithms and polylogarithms, and
in general any cut-free, rational terms are lost.
However, in supersymmetric theories all rational terms
turn out to be uniquely linked to terms with discontinuities,
and therefore the full amplitudes can be reconstructed
in this fashion
\cite{bdk1,bdk2}.

Furthermore, in supersymmetric theories the one-loop
amplitudes are known to be linear combinations
of scalar box functions,
linear triangle functions and linear  bubble functions,
with the coefficients being rational functions
in spinor products.
So the task is really to find an efficient way to fix those
coefficients with as few  manipulations and/or
integrations as possible.

The method based on
conventional unitarity introduced by BDDK
in \cite{bdk1,bdk2} does not evaluate
the phase space integrals
explicitly (from which the full amplitude
would be obtained by performing a dispersion integral),
rather it reconstructs the loop integrand
from which one is able to read off the coefficients
of the various integral  functions.
In practice this means that for a
given momentum channel
the integrand (which is a product of two tree
amplitudes) is simplified as much as possible
using the condition that the two internal lines
are on-shell, and only
in the last step the two delta-functions are replaced
by the appropriate propagators which turn
the integral from a phase space
integral back to a fully-fledged loop integral.
The resulting integral function will have
the correct discontinuities in the particular channel,
but, in general, it will also have
additional discontinuities in other channels.
Nevertheless, working channel by channel one can extract
linear equations for the coefficients
which allow us in the end to determine the complete
amplitude.
However, because of the problem of
the additional, unwanted discontinuities,
this does not provide a diagrammatic method,
i.e.~one cannot just sum the various integrals
for each channel since different discontinuities
might be counted with different weights.

It is natural to contemplate
if there exist other complementary, or more
efficient methods to extract the above mentioned rational
coefficients of the various integral functions,
and if in particular we can replace more than two
propagators by delta functions,  so  that the loop
integration is further restricted - or even
completely localised.
The procedure of replacing several internal
propagators by $\delta^{(+)}$-functions
is well known from the study of
singularities and discontinuities
of Feynman integrals, and goes under the name of
{\it generalised unitarity}  \cite{Cutkosky:1960sp,bible}.
What turns generalised unitarity into a powerful
tool is the fact that generalised cuts of amplitudes
can be evaluated with less effort than
conventional two-particle cuts.

The most dramatic simplification arises from
using quadruple cuts in  one-loop
amplitudes in $\mathcal{N}=4$ SYM.
In this case it is known that the
one-loop amplitudes are simply given by a
sum of scalar box functions without
triangles or bubbles \cite{bdk1}.
Each quadruple cut singles out a unique
box function, and because of the
presence of the four  $\delta^{(+)}$-functions
the loop integration is completely frozen;
hence,  the coefficient of this particular
box is simply given by the product of
four tree-level scattering amplitudes
\cite{bcf-gen}.
An important  subtlety arises here because
quadruple cuts do not have solutions in real Minkowski space;
therefore at intermediate steps one has
to work with complexified momenta.

At this point we can push the analogy
with the ``reconstruction of the
Feynman integrand" a step further.
Using the on-shell conditions we can pull out
the prefactor which is just
the product of four tree-level
amplitudes in front of the integral,
and  the integrand of the remaining
loop integral becomes
just a product of four  $\delta^{(+)}$-functions.
If we now promote the
integral to a Feynman integral
by replacing all $\delta^{(+)}$-functions
by the corresponding propagators%
\footnote{We thank David Kosower
for discussions on this point.}
we arrive at the integral
representation of the appropriate box function.
Note that no overcounting
issue arises, because each
quadruple cut selects a unique box function,
and the final result is
obtained by summing over all
quadruple cuts. In some sense,  one can really think of this
as a true diagrammatic prescription.

As we reduce the amount of supersymmetry
to $\mathcal{N}\!=\!1$,  life becomes a bit more complicated,
since the one-loop amplitudes
are linear combinations of scalar box, triangle and
bubble integral functions.
No ambiguities related to rational terms occur
however, thanks to supersymmetry.
It is therefore natural to attack the problem in two steps:
First, use quadruple cuts to fix all the box coefficients
as described in the previous paragraph.
Second, use triple cuts to fix triangle and
bubble coefficients.
Note that the triple cuts also have contributions
from the box functions which have been determined in
the first step.
The three  $\delta^{(+)}$-functions are not sufficient
to freeze the loop integration completely,
and it is advantageous to use again the
``reconstruction of the Feynman integrand" method,
i.e.~use  the on-shell conditions to simplify
the integrand as much as possible,
and lift the integral to a full loop
integral by reinstating three propagators.
The resulting integrand can be written
as a sum of (integrands of) scalar boxes, triangles
and bubbles, after
standard reduction techniques, like Passarino-Veltman,
have been employed.

At this point it is useful to distinguish
three types of triple cuts according to
the number of external lines attached
to each of the three tree-level amplitudes.
If $p$ of the three amplitudes
have more than one external line
attached, we call the cut a $p$-mass triple cut.
Let us start with the
3-mass triple cut.
The box terms can be dropped as
they have been determined using quadruple cuts,
the coefficients of three-mass triangles
can be read off directly, and the remaining terms,
which are bubbles or triangles with a different
triple cut, are dropped as well.
Special care is needed for 1-mass and 2-mass triple cuts.
First let us  note that any bubble can be written as a linear
combination of scalar and linear 1-mass triangles
or scalar and linear  2-mass triangles depending
on whether the bubble depends on a two-particle invariant,
$t_{i}^{[2]}=(p_{i}+p_{i+1})^{2}$,
or on a $r$-particle invariant,
$t_{i}^{[r]}=(p_{i}+\ldots+p_{i+r-1})^{2}$,
with $r>2$.
Therefore, what we want to  argue is that two-particle
cuts are not needed and that 1-mass, 2-mass {\it and} bubbles
can be  determined from the
1-mass and 2-mass triple cuts.
Now every 1-mass triple cut is in
one-to-one correspondence
with a unique two-particle channel
$t_{i}^{[2]}=(p_{i}+p_{i+1})^{2}$ and
allows us to extract the coefficients of
1-mass triangles and bubbles by only keeping terms in the integral
depending on that particular $t_{i}^{[2]}$
and dropping all boxes and triangles/bubbles
not depending on that  particular variable.
The 2-mass triple cut is associated with
two momentum invariants, say
$P^{2}$ and $Q^{2}$, and we only
keep 2-mass triangles and bubbles
that depend on those two invariants.

In non-supersymmetric theories we have to face the problem that the
amplitudes contain
additional rational terms that are not linked to terms with
discontinuities.  This statement
is true if we only keep terms in the amplitude up to
$\mathcal{O}(\eps^{0})$. If we work
however in $D=4-2 \eps$ dimensions
and keep higher orders in $\eps$, even
rational terms $R$
develop discontinuities of the form
$R (-s)^{-\eps} = R - \eps \log(-s) R + \mathcal{O}(\eps^{2})$
and become cut-constructible%
\footnote{The idea of using unitarity in
$D=4-2 \eps$ dimensions goes back to
\cite{vanNeerven:1985xr}, and was used in
\cite{Bern:1995db,Bern:1996ja}. }.
In practice, this means that,  in our procedure,
whenever we cut internal lines by replacing
propagators by $\delta^{(+)}$-functions
we have to keep the cut lines in $D$ dimensions,
and in order  to proceed we need to know tree-amplitudes with
two legs continued to $D$ dimensions.
Because of the supersymmetric decomposition of one-loop
amplitudes in pure Yang-Mills, which was reviewed in
the Introduction, we only
need to consider the case
of a scalar running in the loop.
Furthermore, the massless scalar in $D$
dimensions can be thought of
as a massive scalar in four dimensions
$L^{2}=l^{2}_{(4)}+l^{2}_{(-2\eps)}=l^{2}_{(4)}\!-\!\mu^{2}$
whose mass has to be integrated over \cite{Bern:1995db,Bern:1996ja}.
Interestingly, a term in the loop integral with the insertion of
``mass'' term   $(\mu^{2})^{m}$
can be mapped to a higher-dimensional loop integral
in $4+2m-2\eps$  dimensions with a massless scalar
\cite{Bern:1995db,Bern:1996ja}.
Some of the required tree amplitudes with
two massive scalars and all
positive helicity gluons
have been calculated in
\cite{Bern:1995db,Bern:1996ja}
using Feynman diagrams and recursive techniques,
and more recently all amplitudes
with up to four arbitrary helicity gluons and
two massive scalars have been presented in \cite{snvp}.

The comments in the last paragraph make it clear that generalised
unitarity techniques can readily be generalised
to $D$ dimensions and be used to obtain
complete amplitudes in pure Yang-Mills
and, more generally, in massless,
non-supersymmetric gauge theories.
The integrands produced by the method described for four dimensional
unitarity will now
contain terms multiplied by $(\mu^{2})^{m}$
and, therefore, the set of integral functions
appearing in the amplitudes
includes,  in addition to the
four-dimensional functions,  also higher-dimensional
box, triangle and bubble functions (some explicit examples of
higher-dimensional integral functions
can be found in Appendix A). For example the
one-loop $+$$+$$+$$+$
gluon amplitude, which vanishes in SYM,
is given by a rational function
times a box integral with $\mu^{4}$
inserted, $I_{4}[\mu^{4}]=(-\eps)(1-\eps)
I_{4}^{8-2\eps}=-1/6+\mathcal{O}(\eps)$.
Hence this amplitude
is a purely rational function in spinor variables.

In the following sections we will describe in detail
how this procedure is applied in practice
by recalculating  all four-gluon scattering amplitudes
and the positive helicity five-gluon scattering amplitude
in pure Yang-Mills at one-loop level.
These examples include the cases of infrared finite
amplitudes that are purely rational
(and their supersymmetric counterparts
vanish), and infrared divergent amplitudes that
contain both rational and cut-constructible terms.


\section{Four-point amplitudes in pure Yang-Mills}


In this section we recalculate all the known four-gluon
scattering amplitudes, that is
$+$$+$$+$$+$, $-$$+$$+$$+$, $-$$-$$+$$+$, and finally $-$$+$$-$$+$,
from quadruple and triple cuts.

\subsection{The one-loop $+$$+$$+$$+$ amplitude}
The one-loop $+$$+$$+$$+$ amplitude with a complex scalar running in the loop
is the simplest of the all-plus gluon amplitudes,
and was first derived in \cite{bk}
using the string-inspired formalism.

The expression in $4-2\e$ dimensions, valid to all-orders in $\e$,
is computed in
\cite{Bern:1995db}
and is given by
\begin{equation} \label{++++}
 \cA_4^{\rm scalar}(1^+,2^+, 3^+, 4^+) \ = \
  \frac{2i}{(4\pi)^{2-\epsilon} }
\frac{[12][34]}{\lan 12\ran\lan 34\ran }
  \ K_4 \ ,
\end{equation}
where%
\footnote{Notice also that
$[12] [34] / (\lan 12 \ran \lan 34 \ran ) = - s_{12} s_{23}/
(\lan 12 \ran \lan 23 \ran \lan 34 \ran  \lan 41 \ran )$.
}
\begin{equation}
\label{K4}
K_4 \ := \ I_4[\mu^4] \ = \
-\eps(1-\eps)I_4^{D=8-2\eps} \ = \ -\frac{1}{6} + \cO(\eps)
\ .
\end{equation}
In this paper we closely follow
the conventions of \cite{Bern:1995db},
with 
\beq
\label{integral}
I_n^{D=4-2\e} [ f(p, \mu^2) ]   \ := \
i (-)^{n+1} (4 \pi )^{2-\e}
\int \!\! {d^4 l \over (2 \pi)^4} {d^{-2\e}\m \over (2 \pi)^{-2\e}}
{ f(l, \mu^2)  \over (l^2 - \m^2)
\cdots [ (l - \sum_{i=i}^{n-1} K_i)^2 - \m^2]}
\ ,
\eeq
where $K_i$ are external momenta
(which, in colour-ordered amplitudes,
are sums of adjacent null momenta of the external gluons)
and $f(l , \mu^2)$
is a generic function of the four-dimensional loop momentum
$l$ and of $\mu^2$.

The amplitude with four positive helicity gluons
is part of the infinite sequence of
all-plus helicity gluons, for which a closed
expression was conjectured  in \cite{Bern:1993sx,Bern:1993qk}.
The result for all $n$ is given by
\beq
\cA_n  (+,\dots,+)\ =\ - {i \over 48 \pi^2}
\sum_{1\leq i_1<i_2<i_3<i_4\leq n}
{ \lan i_1 i_2\ran  \, [i_2 i_3] \, \lan i_3 i_4\ran
\, [i_4 i_1] \over \lan  1 2 \ran   \, \lan 2 3 \ran
\cdots  \lan n 1\ran }
\ ,
\eeq
or, alternatively,
\beq
\cA_n \ = \ - {i \over 96 \pi^2}
\sum_{1\leq i_1<i_2<i_3<i_4\leq n} { s_{i_1 i_2} s_{i_3i_4}
\, - \, s_{i_1i_3}s_{i_2i_4}\, + \, s_{i_1i_4}s_{i_2i_3}\, -\,
4i
\epsilon (i_1 i_2 i_3 i_4)
\over \lan  1 2\ran  \, \lan 2 3 \ran \cdots
\lan n 1\ran }
\ ,
\eeq
where $\e (abcd) := \e_{\m \n \r s} a^\m b^\n c^\r d^\s$.
As $\e \to 0$, \eqref{++++} becomes
\beq
\cA_4 \ = \ {i \over 48 \pi^2} {s_{12} s_{23} \over
\lan 12 \ran \lan 23  \ran \lan 34 \ran\lan 41 \ran
 }
\ .
\eeq
We see that this amplitude \eqref{++++}
consists of purely rational terms,
which are cut-free in four dimensions.
We now show how to derive \eqref{++++}
from quadruple cuts in $D\!=\! 4-2\e$ dimensions.

\begin{figure}[ht]
\label{figure-quad1}
\begin{center}
\scalebox{0.65}{\includegraphics{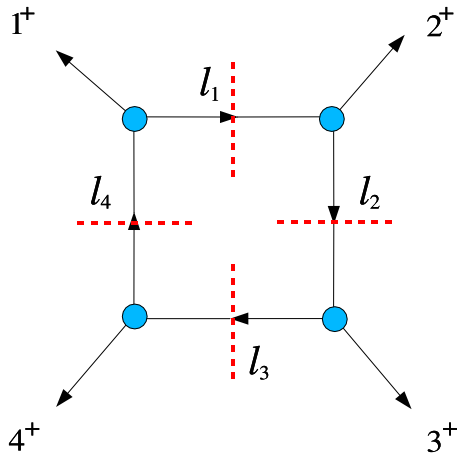}}
\end{center}
\caption{\it
One of the two quadruple-cut diagrams
for the amplitude $1^+ 2^+ 3^+ 4^+$.
This diagrams is obtained by sewing
tree amplitudes (represented by the blue bubbles)
with an external positive-helicity gluon and two
internal scalars of opposite ``helicities''.
There are two such diagrams, which are obtained
one from the other by flipping all the internal
helicities. These diagrams are equal
so that the full result is obtained by doubling
the contribution from the diagram in this Figure.
The same remark applies to all the other diagrams
considered in this paper.
}
\end{figure}

\begin{figure}[ht]
\label{figure-triangle1}
\begin{center}
\scalebox{0.70}{\includegraphics{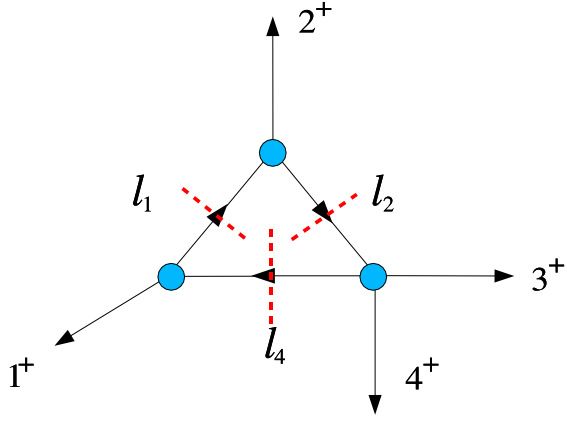}}
\end{center}
\caption{\it
One of the possible three-particle cut diagrams
for the amplitude  $1^+ 2^+ 3^+ 4^+$.
The others are obtained from this one by 
cyclic relabeling of the external particles.
}
\end{figure}

Consider the quadruple-cut diagram in Figure 1,
which is obtained by sewing
four three-point scattering amplitudes%
\footnote{In the following for the purpose of calculating the (generalised) cuts we
drop factors of $i$ appearing in the usual definition of tree-amplitudes and propagators.
For quadruple and two-particle cuts this does not affect the final result, while for triple cuts
this introduces an extra $(-1)$ factor which we reinstate at the end of every calculation.}
with one
massless gluon and two massive scalars
of mass $\mu^2$. From \cite{snvp}
we take the three-point amplitudes
for one positive-helicity gluon and two scalars:
\beq
\label{threepoint}
\cA (l_1^+ , k^+ , l_2^- ) \ = \
\cA (l_1^- , k^+ , l_2^+ ) \ = \
{\lan q  | l_1 | k ] \over \lan q  k \ran}
\ ,
\eeq
where $l_1 + l_2 + k =0$. Here $|q \ran$
is an arbitrary reference spinor not proportional
to $|k \ran$. It is easy to see  \cite{snvp}
that \eqref{threepoint} is actually
independent of the choice of  $|q \ran$.

The $D$-dimensional quadruple
cut of the amplitude $+$$+$$+$$+$
is obtained by combining four three-point
tree-level amplitudes,
\begin{equation} \label{4vertices}
\frac{\lan q_1\vert l_1 \vert 1]}{\lan q_11\ran }\
\frac{\lan q_2\vert l_2 \vert 2]}{\lan q_22\ran }\
\frac{\lan q_3\vert l_3 \vert 3]}{\lan q_33\ran }\
\frac{\lan q_4\vert l_4 \vert 4]}{\lan q_44\ran }
\ .
\end{equation}
The reference momenta $q_i, i=1, \ldots ,4$ in each of
the four ratios in this expression
may be chosen arbitrarily.
Then, using momentum conservation,
\beq
l_2 = l_1 - k_2 \ , \qquad
l_4 = l_3 - k_4 \ ,
\eeq
the fact that the external
momenta are null,
and that the internal momenta square to $\mu^2$,
it is easy to see that
\beq
\label{simplific}
\frac{
\lan q_1\vert l_1 \vert 1]}{\lan q_1 1\ran }\
\frac{\lan q_2\vert l_2 \vert 2]}{\lan q_2 2\ran }\
 = - \m^2 {[12] \over \lan 12 \ran }
\ ,
\eeq
and similarly
\beq
\frac{\lan q_3\vert l_3 \vert 3] }{\lan q_3 3\ran }\
\frac{\lan q_4\vert l_4 \vert 4]}{\lan q_4 4\ran }\
 = - \m^2 { [34 ] \over \lan 34 \ran}
\ ,
\eeq
so that the above expression \eqref{4vertices}
becomes simply
\begin{equation}
\label{4vertices2}
\mu^4 \, \frac{[12][34]}{\lan 12\ran\lan 34\ran}
\ .
\end{equation}
Finally, we lift the quadruple-cut box to a
box function by reinstating the appropriate
Feynman propagators.
These propagators then combine with the
additional factor of $\mu^4$ in \eqref{4vertices2}
to yield the factor $ i K_4/ {(4\pi)^{2-\eps}}$ 
which is proportional to the scalar box integral 
defined in \eqref{K4}. Including an additional 
factor of 2 due to the fact that there is a complex
scalar propagating in the loop, 
the amplitude \eqref{++++} is reproduced correctly.

Next we inspect three-particle cuts.
One of the three tree-level amplitudes we sew in the triple-cut amplitude
is an amplitude with two positive-helicity gluons and two scalars \cite{Bern:1996ja}
\beq
\label{fourpoint}
\cA ( l_1^+ , 1^+ , 2^+, l_2^-)
\ = \ \m^2 {[12] \over \lan 12 \ran [ (l_1 + k_1)^2 - \m^2 ] }
\ .
\eeq
Consider, for example, the three-particle cut
defined by $1^+, 2^+, (3^+, 4^+)$, see Figure 2.
Using \eqref{threepoint} and \eqref{fourpoint},
the product of the three tree-level amplitudes gives
\beq
\frac{\lan q_1\vert l_1 \vert 1]}{\lan q_11\ran }\
\frac{\lan q_2\vert l_1 \vert 2]}{\lan q_22\ran }\
 { \m^2 [34] \over \lan 34  \ran [ (l_2 - k_3)^2 - \m^2 ] }
\ ,
\end{equation}
with $l_2 = l_1 - k_2$.
As for the quadruple cut, it is easily seen that,
on this triple cut,
\beq
\frac{\lan q_1\vert l_1 \vert 1]}{\lan q_11\ran }\
\frac{\lan q_2\vert l_1 \vert 2]}{\lan q_22\ran }
\ = \ - \m^2 {[12] \over \lan 12 \ran}
\ ,
\eeq
where we used $l_1^2 =  l_2^2 = l_4^2 =  \m^2$.
The triple-cut integrand then becomes
\beq
- {[12][34] \over \lan 12 \ran\lan 34\ran}
{\m^4 \over [ (l_2 - k_3)^2 - \m^2 ]}
\ ,
\eeq
which, after replacing the three $\d^{(+)}$ functions 
by propagators, 
integrates to 
\eqref{++++},   
where we have included an additional $(-1)$ 
factor following the comments in footnote 6.
The factor of $2$ in \eqref{++++} comes from summing over
the two ``scalar helicities''. 
The same result  comes from 
evaluating the remaining triple cuts.

We remark that in the case
of the quadruple cut we did not even need
to insert the solutions of the on-shell 
conditions for the loop momenta
into the expression coming from the cut. This is not true in general;
for example, for the five gluon amplitude discussed below the
sum over solutions will be essential to obtaining the correct amplitude.


\subsection{The one-loop $-$$+$$+$$+$ amplitude}
The one-loop four gluon scattering amplitude $-$$+$$+$$+$, with a complex 
scalar running in the loop, is given
to all orders in $\e$ by \cite{Bern:1995db}
\begin{eqnarray} \label{+++-}\nonumber
\cA_4^{\rm scalar}(1^-,2^+, 3^+, 4^+) &  = &
  \frac{2 i}{(4\pi)^{2-\eps}}
  { [24]^2 \over [12] \lan 23\ran  \lan 34\ran [41]}
  { s t \over u } \,
  \biggl[
      { t(u-s) \over s u } J_3(s)
      + { s(u-t) \over t u } J_3(t)
 \\  \nonumber
            & -&  { t-u \over s^2 } J_2(s)
      - { s-u \over t^2 } J_2(t)
      + { s t \over 2 u } J_4
      + K_4 \biggr]
\ .
\\
\end{eqnarray}
We will now show how to derive this
result using generalised unitarity cuts.

First consider the quadruple cut (see Figure 3).
\begin{figure}[ht]
\label{figure-quad3}
\begin{center}
\scalebox{0.65}{\includegraphics{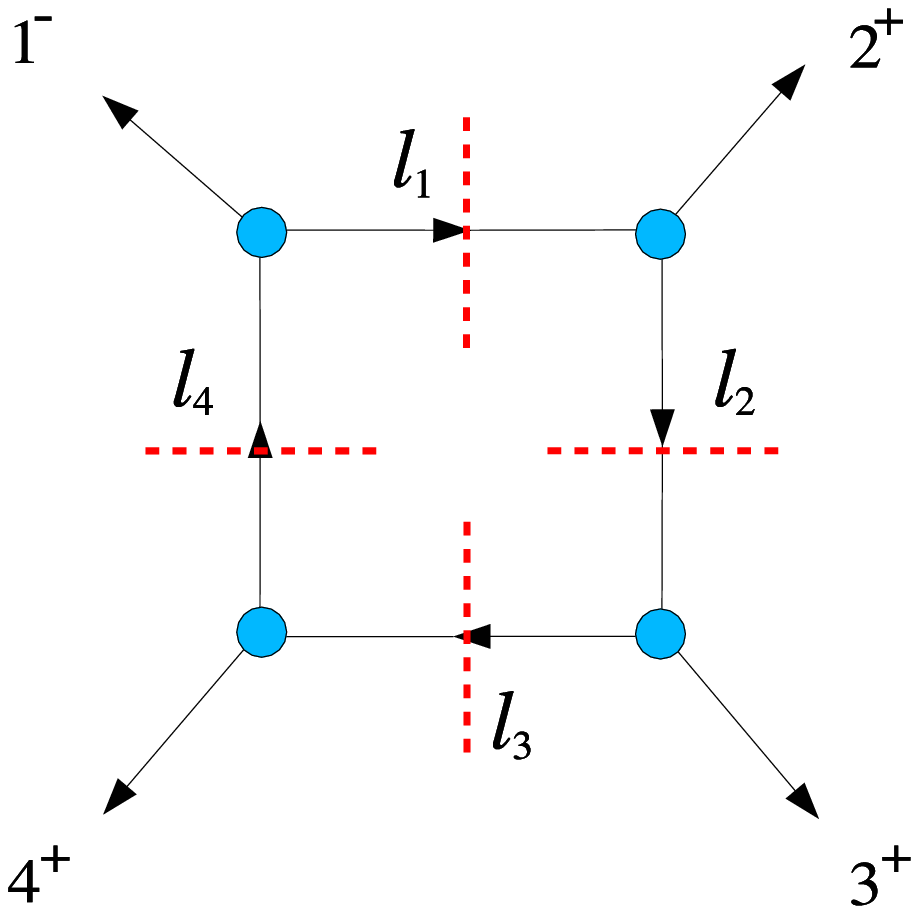}}
\end{center}
\caption{\it
The quadruple cut
for the amplitude $1^- 2^+ 3^+ 4^+$.
}
\end{figure}
The product of tree amplitudes gives
\begin{equation}
\frac{\lan 1\vert l_1 \vert q_1]}{[1q_1]}\
\frac{\lan q_2\vert l_2 \vert 2]}{\lan q_22\ran }\
\frac{\lan q_3\vert l_3 \vert 3]}{\lan q_33\ran }\
\frac{\lan q_4\vert l_4 \vert 4]}{\lan q_44\ran }\ .
\end{equation}
It is straightforward to show that, on the
quadruple cut,
\begin{eqnarray}
\label{interm1}\nonumber
\frac{\lan q_3\vert l_3 \vert 3]}{\lan q_33\ran }\
\frac{\lan q_4\vert l_4 \vert 4]}{\lan q_44\ran }
&
=& - \mu^2\frac{[34]}{\lan 34\ran}\ ,
\\  \nonumber
\frac{\lan 1\vert l_1 \vert q_1]}{\lan 1q_1\ran }\
\frac{\lan q_2\vert l_2 \vert 2]}{\lan q_22\ran }
&=& \frac{[23]}{[31]}\bigg( -\mu^2\frac{\lan 31\ran}{\lan 23\ran}
  \, - \,  [2\vert l_1 \vert 1\ran \bigg)\ ,
\end{eqnarray}
and hence the quadruple cut in Figure 3 gives
\beq\label{quadruple}
Q(1^+,2^+,3^+,4^-) =  \mu^2 {[3\, 4]\over \lan 3 \, 4\ran } \,
{[2\, 3] \over [3 \, 1] }
\left[ \mu^2 {\lan 3\,  1 \ran  \over \lan 2 \, 3\ran } \ +
\ [2 |l_1 |1 \ran
\right]
\ .
\eeq
In order to compare with \eqref{+++-} it is useful to notice that
\beq
\label{enne}
{[34]\over \lan 34\ran  } {[23]\over  [31] }
{\lan 31 \ran \over \lan 23 \ran } \ = \
 { [24]^2 \over [12] \lan 23\ran  \lan 34\ran [41]}
  { s t \over u } \ := \
\cN
\ .
\eeq
We conclude that the first term in \eqref{quadruple}
generates
\beq
\label{primo}
\frac{i}{(4\pi)^{2-\eps}} \,
\left( { [24]^2 \over [12] \lan 23\ran  \lan 34\ran [41]}
  { s t \over u } \right)
\ K_4
\ ,
\eeq
where the prefactor  in \eqref{primo} comes from
the definition \eqref{K4} and 
\eqref{integral} for the function $K_4$.

The second term in \eqref{quadruple}
corresponds to a linear box integral, which we examine now.
We notice that the quadruple cut freezes
the loop integration on the solution for the cut.
In the linear box term in \eqref{quadruple}
we will then replace $l_1$ in $[2 |l_1 |1 \ran$
by the solutions of
the cut, and sum  over the different solutions.

Specifically, in order to solve for the cut-loop momentum $l_1$
one has to require
\begin{eqnarray}\label{momcon}\nonumber
&&l_1^2\, =\, l_2^2\, = \, l_3^2\, = \, l_4^2\, = \, \mu^2 \ ,
\\ [6pt]
&&l_1\, =\, l_4-k_1 \ , \
l_2\, =\, l_1-k_2\ , \
l_3\, = \, l_2-k_3\ , \
l_4\, =\, l_3-k_4\ .
\end{eqnarray}
%
In order to solve these conditions, it proves useful
\cite{bcf-gen} to use 
the four linearly independent
vectors $k_1,k_2,k_3$ and $K$, where
\beq\label{Kthingy}
K_\mu\ := \ \epsilon_{\mu\nu \rho\sigma}\, k_1^\nu
\,  k_2^\rho \, k_3^\sigma
\ .
\eeq
Setting
\beq
\label{elvira}
l_1 \  = \ ak_1\, +\, bk_2\, +\, ck_3\, +\, dK \ ,
\eeq
one finds
\begin{eqnarray}
\label{dongiovanni}
&&a  \, =\, \frac{t}{2u}\, , \ b\, =\, \frac{1}{2}\, ,\
c\, =\, -\frac{s}{2u}\, ,
\\ [6pt]\nonumber
&&d \, =\, \pm \sqrt{-{st+4\mu^2u\over s t u^2}} \, ,
\eeqa
where
\beq
\label{stu}
s\, = \, (k_1+k_2)^2\  ,\ t \, =\,  (k_2+k_3)^2 \ ,\ u
\, = \,  (k_1+k_3)^2
\ ,
\eeq
and $s + t + u =0$.

Then one has
\beq
[2 |l_1 |1 \ran \longrightarrow [2 |{l_1^+ + l_1^-\over 2} |1 \ran
 \ = \
 c \cdot [2 |3 |1 \ran \ = \
- {s \over 2u} [23] \lan 31 \ran
\ ,
\eeq
where $l_1^\pm$ denotes the two solutions for the quadruple cut.
The square root drops out of the calculation (as it should,
given that the amplitude is a rational function).
We conclude that the second term in \eqref{quadruple}
gives%
\footnote{Recall that in our conventions
$t := \lan 23 \ran [32] $.}
\beq
\label{pluto}
\frac{i}{(4\pi)^{2-\eps}} \,  
\left( { [24]^2 \over [12] \lan 23\ran  \lan 34\ran [41]}
  { s t \over u } \right)
\ {st \over 2u} J_4
\ ,
\eeq
where
\beq
J_n \ :=\ I_n [ \mu^2]
\  .
\eeq
Again, the prefactor in \eqref{pluto} arises from the definition
\eqref{integral}. 


In total the quadruple cut \eqref{quadruple} gives
\beq
\frac{2 i}{(4\pi)^{2-\eps}} \, \cN \left( K_4 \ + \ {st \over 2u} J_4 \right)
\ ,
\eeq
where we have again 
included a factor of two for 
the contribution of a complex scalar. This result matches
exactly all the box functions appearing in \eqref{+++-}.

\begin{figure}[ht]
\label{Figure3}
\begin{center}
\scalebox{0.70}{\includegraphics{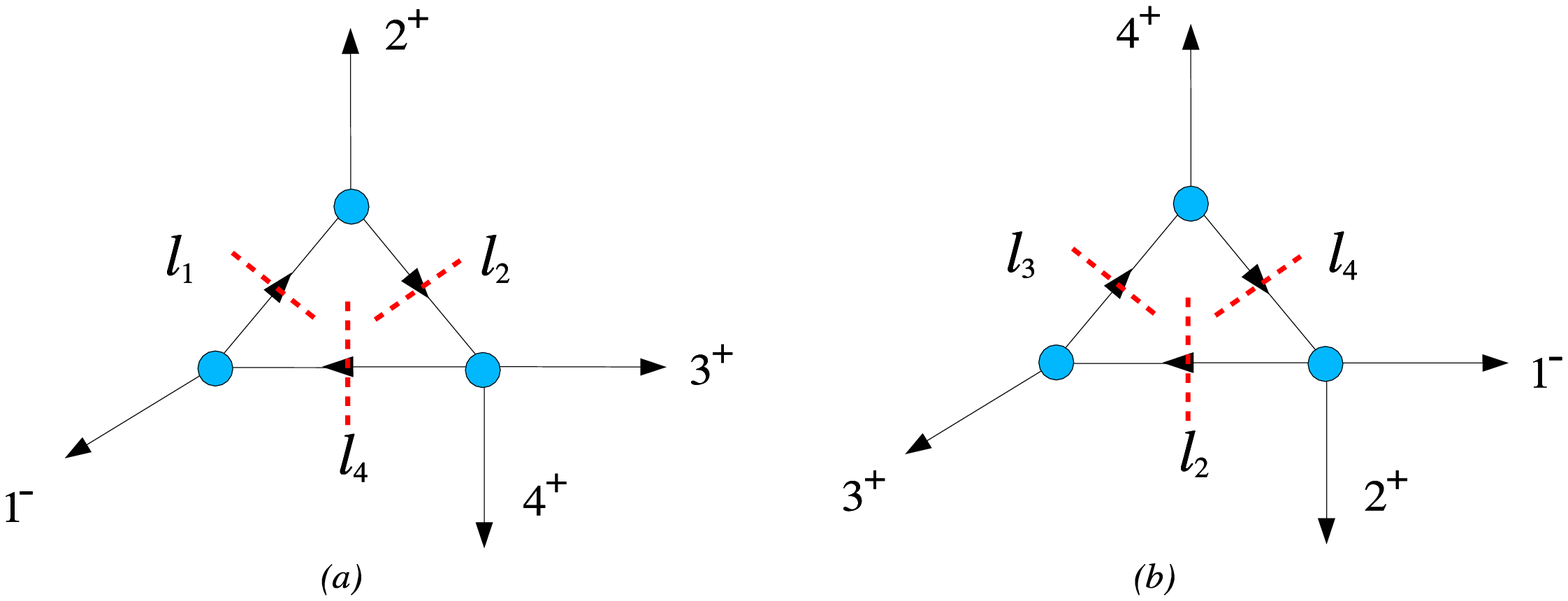}}
\end{center}
\caption{\it
The two inequivalent triple cuts
for the amplitude  $1^- 2^+ 3^+ 4^+$.
}
\end{figure}

We now move on to consider triple cuts.
We start by considering the triple cut
in Figure 4{\it a}, which we label as
$(1^- , 2^+ , (3^+ , 4^+))$.
It may be shown that this triple cut yields
the following expression:
\beqa\label{triple1.2.34}\nonumber
TC(1^- , 2^+ , (3^+ , 4^+)) & = &
\mu^2 {[3\, 4]\over \lan 3 \, 4\ran } \,
{[2\, 3] \over [3 \, 1] }
\left( -\mu^2 {\lan 3\,  1 \ran  \over \lan 2 \, 3\ran } \ -
\ [2 |l_1 |1 \ran
\right) {1 \over (l_2 - k_3)^2 - \mu^2} \
\\ [8pt] \qquad\qquad
&& \qquad -   \mu^2 \,
{[3 \, 4] \over \lan 3 \, 4 \ran }
{[2 |l_1 |1 \ran\over \lan 2 \, 3 \ran [3 \, 1 ]
 }
\ .
\eeqa
The first line in \eqref{triple1.2.34} clearly contains
the (negative of the) term
already studied with quadruple cuts --
see \eqref{quadruple} (for an explanation of 
the relative minus sign see footnote 6). 
We now reconsider the
linear box term
(second term in the first line of
\eqref{triple1.2.34}),
and study its Passarino-Veltman (PV) reduction.
As we shall see, this box appears also in other triple cuts
(see \eqref{triple1.2.34-reloaded}).

Let us consider the linear box integral
\beq
A^\m \ := \
\int \!\! {d^4 l_1
\over (2 \pi)^4} {d^{-2\e}\m \over (2 \pi)^{-2\e}}
{\m^2 \  l_1^\m\over (l_1^2 - \m^2) [ (l_1 - k_2)^2 - \m^2]
[ (l_1 - k_2-k_3)^2 - \m^2] [(l_1 + k_1 )^2 - \m^2]
}
\ .
\eeq
On general grounds 
the integral is a linear combination 
of three of the external momenta,
\beq
A^\m \ = \ \a k_1^\m \, + \, \b k_2^\m  + \, \g k_3^\m
\ .
\eeq
For the coefficients we find
\beqa
\a & = & 
- { i \over (4\pi)^{ 2 - \e}} 
\ 
{1\over 2u} \Big[ -t J_4 \, - \,
2 J_3 (s) \, + \, 2 J_3 (t) \Big]
\ ,
\\ \nonumber
\b & = & 
 { i \over (4\pi)^{ 2 - \e}} 
\ 
{1\over 2} J_4
\ ,
\\ \nonumber
\g & = & 
- { i \over (4\pi)^{ 2 - \e}} 
\ 
{1\over 2u} \Big[ s J_4 \, - \, 2 J_3 (s) \, + \, 2 J_3 (t)
\Big]
\ . 
\eeqa
Taken literally, this means that from the linear box in
 \eqref{triple1.2.34}  we not only get the $J_4$ function
but, altogether:
\beq
\, \frac{i \cN}{(4 \pi)^{2-\eps}} \left(
{st \over 2u} J_4  \, - \, {t\over u} J_3 (s) \, + \,
{t \over u} J_3 (t) \right)
\ .
\eeq
Summarising, the PV reduction of the first line of the triple cut
 \eqref{triple1.2.34},  lifted to
a Feynman integral,  gives:
\beq
\label{PVquadruplelifted}
\frac{i \cN}{(4 \pi)^{2-\eps}} \left(
K_4 \, + \, {st \over 2u} J_4  \, - \, {t\over u} J_3 (s) \, + \,
{t \over u} J_3 (t) \right)
\ .
\eeq
The last term
in \eqref{PVquadruplelifted} is clearly spurious --
it does not have the right
triple cut, and has appeared because we
lifted  the cut-integral
to a Feynman integral; hence we will drop it.
In conclusion, the triple cut $(1^- , 2^+ , (3^+ , 4^+))$
in Figure 4{\it a} leads to
\beq
\label{PVtriple1}
\frac{i \cN}{(4 \pi)^{2-\eps}} \left(
K_4 \, + \, {st \over 2u} J_4  \, - \, {t\over u} J_3 (s)
\right)
\ .
\eeq
We now consider the last term in  \eqref{triple1.2.34},
which  generates a linear triangle, whose
PV reduction we consider now.
The linear triangle is proportional to
\beq
B^\m \ := \
\int \!\! {d^4 l_1 \over (2 \pi)^4} {d^{-2\e}\m \over (2 \pi)^{-2\e}}
{ \m^2 \, l_1^\m\over (l_1^2 - \m^2) [ (l_1 - k_2)^2 - \m^2]
[ (l_1 + k_1 )^2 - \m^2]
}
\ .
\eeq
On general grounds,
\beq
B^\m \ = \ \theta k_1^\m \, + \, \tau k_2^\m
\ , 
\eeq
and hence
\beq
[2 |\, B \,  |1 \ran\ \ = \ 0
\ .
\eeq
We conclude that the second line in
\eqref{triple1.2.34} gives a vanishing contribution, so that the
content of this triple cut is encoded in
\eqref{PVtriple1}.

Next we consider the  triple cut
labelled by $((1^- , 2^+) , 3^+ , 4^+)$ and
represented in Figure 4{\it b},
which gives
\beqa\label{triple12.3.4}\nonumber
TC((1^- , 2^+) , 3^+ , 4^+) & = &
\mu^2 {[3\, 4]\over \lan 3 \, 4\ran }
{[2\, 3] \over [3 \, 1] }
\left[ -\mu^2 {\lan 3\, 1 \ran  \over \lan 2 \, 3\ran  }
\ + \
{\lan 1 \, 2\ran  \over\lan 2 \, 3\ran }\,
\lan 3 |l_2 |2 ]
\right]{1 \over (l_2 +  k_2)^2 - \mu^2} \
\\ [8pt]
&&\qquad +
\mu^2 {[3\, 4]\over \lan 3 \, 4\ran } \,
{\lan 1 | 3\, 1\, l_2 - 2\, 3 \, l_2 |2 ] \over
\lan 1 \, 2 \, \ran [1 \, 2] \, \lan 2 \, 3 \ran \, [3 \, 1]}
\ .
\eeqa
The first term of
\eqref{triple12.3.4}
clearly corresponds to
the function $K_4$ already fixed using quadruple cuts.
The second term can be rewritten as follows.
Introducing $l_1 := l_2 + k_2$,
we have
\beq
 {\lan 12 \ran \over \lan 23  \ran }\lan 3 | l_2| 2] \ = \
- \, [2 |l_2 | 1\ran \,  + \, {\lan 13 \ran \over \lan 23 \ran} \,
 [(l_2 +  k_2)^2 - \m^2]
\ ,
\eeq
therefore we can rewrite \eqref{triple12.3.4}
as
\beqa\label{triple1.2.34-reloaded}\nonumber
TC((1^- , 2^+) , 3^+ , 4^+) & = &
\mu^2 {[3\, 4]\over \lan 3 \, 4\ran } \,
{[2\, 3] \over [3 \, 1] }
\left( -\mu^2 {\lan 3\,  1 \ran  \over \lan 2 \, 3\ran } \ -
\ [2 |l_1 |1 \ran
\right) {1 \over (l_2 + k_2)^2 - \mu^2} \
\\ [8pt] \qquad\qquad
& + &
\mu^2 {[3\, 4]\over \lan 3 \, 4\ran }
\,
\biggl(
{\lan 1 | 3\, 1\, l_2 - 2\, 3 \, l_2 |2 ] \over
\lan 1 \, 2 \, \ran [1 \, 2] \, \lan 2 \, 3 \ran \, [3 \, 1]}
 \, - \, {[23] \over [31]} {\lan 31 \ran \over \lan 23 \ran}
\biggr)
\ .
\eeqa
We know already that the  PV reduction of the first line of
\eqref{triple1.2.34-reloaded} corresponds to
\eqref{PVquadruplelifted}
-- with the term containing $J_3 (t) $
removed --
so we now study the second line, which will give
new contributions.

The second term in the second line corresponds
to a scalar triangle,
more precisely it gives a contribution
\beq
- \,\frac{i \cN}{(4 \pi)^{2-\eps}} \,  J_3 (s)
\ .
\eeq
The first term 
corresponds to a linear triangle, and now we
perform its PV reduction.
The relevant integral is
\beq
C^\m \ := \
\int \!\! {d^4 l_2 \over (2 \pi)^4} {d^{-2\e}\m \over (2 \pi)^{-2\e}}
{ \m^2 \, l_2^\m\over (l_2^2 - \m^2) [ (l_2 - k_3)^2 - \m^2]
[ (l_2 + k_1 + k_2 )^2 - \m^2]
}
\ .
\eeq
On general grounds,
\beq
C^\m \ = \ \l \, k_3^\m \ + \ \kappa \, (k_1 + k_2)^\m
\ .
\eeq
A quick calculation shows that
\beq
\l \ = 
- {i \over (4 \pi)^{2 - \e}}
\  
\Bigl[ 
J_3 (s) \, -  \, {2\over s} J_2 (s)
\Bigr] 
\ , \qquad
\k \ = \  {i \over (4 \pi)^{2 - \e}}  \ 
{1\over s}  J_2 (s)
\ .
\eeq
The first term  in the second line of
\eqref{triple1.2.34-reloaded} gives then
\beq
\frac{i \cN}{(4 \pi)^{2-\eps}} \ 
\left(-  {u\over s} \, J_3 (s) \, +\,  {u-t \over s} \,
J_2 (s)
 \right)
\ ,
\eeq
where $\cN$ is defined in \eqref{enne}.
Altogether, the second line of \eqref{triple1.2.34-reloaded}
gives
\beq
\label{bb}
\frac{i \cN}{(4 \pi)^{2-\eps}} \ \left(-  \Big( 1 \, + {u\over s} \Big)  \, J_3 (s) \, + \,
{u-t \over s} \,
J_2 (s)
 \right)
\ ,
\eeq
whereas from the first line of the same equation
we get 
\beq
\label{PVquadruplelifted3}
\frac{i \cN}{(4 \pi)^{2-\eps}} \left(
K_4 \, + \, {st \over 2u} J_4  \, - \, {t\over u} J_3 (s)
\right)
\ ,
\eeq
where we have dropped the term $J_3 (t)$ for reasons explained
earlier.

We conclude that the function
which incorporates all the right cuts
in the channels considered so far is equal to
the sum of \eqref{bb} and \eqref{PVquadruplelifted3},
which gives
\beq
\label{above}
\frac{i \cN}{(4 \pi)^{2-\eps}} \left( K_4 \, + \, {st \over 2u} J_4
\, - \, {t \over u}  \, J_3 (s) \, - \,
\Big( 1 \, + {u\over s} \Big)  \, J_3 (s) \, + \,
{u-t \over s^2} \, J_2 (s) \right)
\ .
\eeq
Using $-t/u - 1 - u/s = s/u - u/s$,  \eqref{above} becomes
\beq
\frac{i \cN}{(4 \pi)^{2-\eps}} \left( K_4 \, + \, {st \over 2u} J_4
\, +  \,
\Big( {s \over u}  \, - \,  {u\over s} \Big)  \, J_3 (s) \, + \,
{u-t \over s^2} \, J_2 (s) \right)
\ .
\eeq
To finish the calculation one has to consider
the two remaining triple  cuts, that is
$(4^+ , 1^- , (2^+ , 3^+))$
and  $((4^+ , 1^-) , 2^+ , 3^+)$.
These cuts can be obtained
from the previously considered cuts by
exchanging $s$ with  $t$.

Our conclusion is therefore that the function (including the usual factor of 2)
with the  correct quadruple and triple cuts is:
\beqa
&&\frac{2 i \cN}{(4 \pi)^{2-\eps}} \left(
 K_4 \, + \, {st \over 2u} J_4
+ 
\Big( {s \over u}  \, - \,  {u\over s} \Big)  \, J_3 (s) \, + \,
{u-t \over s^2} \, J_2 (s) \right.
\\ \nonumber
& & \qquad \qquad\qquad \qquad + \left.  \Big( {t \over u}  \, - \,  {u\over t} \Big)  \, J_3 (t) \,
+ \,
{u-s \over t^2} \, J_2 (t) \right)
\ .
\eeqa
This agrees precisely with \eqref{++++} using the identities
\beq
{t (u-s) \over su} \ = \ {s \over u}  \, - \,  {u\over s}
\  , \qquad
{s (u-t) \over tu} \ = \ {t \over u}  \, - \,  {u\over t}
\ .
\eeq

\subsection{The one-loop $-$$-$$+$$+$ amplitude}

We now turn our attention to the one-loop four point
amplitudes with two negative helicity gluons.
We start by considering the one-loop amplitude $\cA^{\rm scalar}_4(1^-,2^-,3^+,4^+)$,
which is given by  \cite{Bern:1995db}%
\footnote{Here for simplicity we drop the functions 
$I_1$ and $I_2(0)$,
which are zero in the massless case \cite{Bern:1995db}. We also include a factor of two as we are considering complex scalars.}
\beq\label{4pt--++}
\cA^{\rm scalar}_4(1^-,2^-,3^+,4^+) = 2\frac{\cA_4^{\rm tree}}{(4\pi)^{2-\epsilon}}
\bigg( -\frac{t}{s}K_4 + \frac{1}{s}J_2(t) +
\frac{1}{t}I_2^{6-2\epsilon}(t)
\bigg)\ .
\eeq
\begin{figure}[ht]
\label{figure-quad4}
\begin{center}
\scalebox{0.65}{\includegraphics{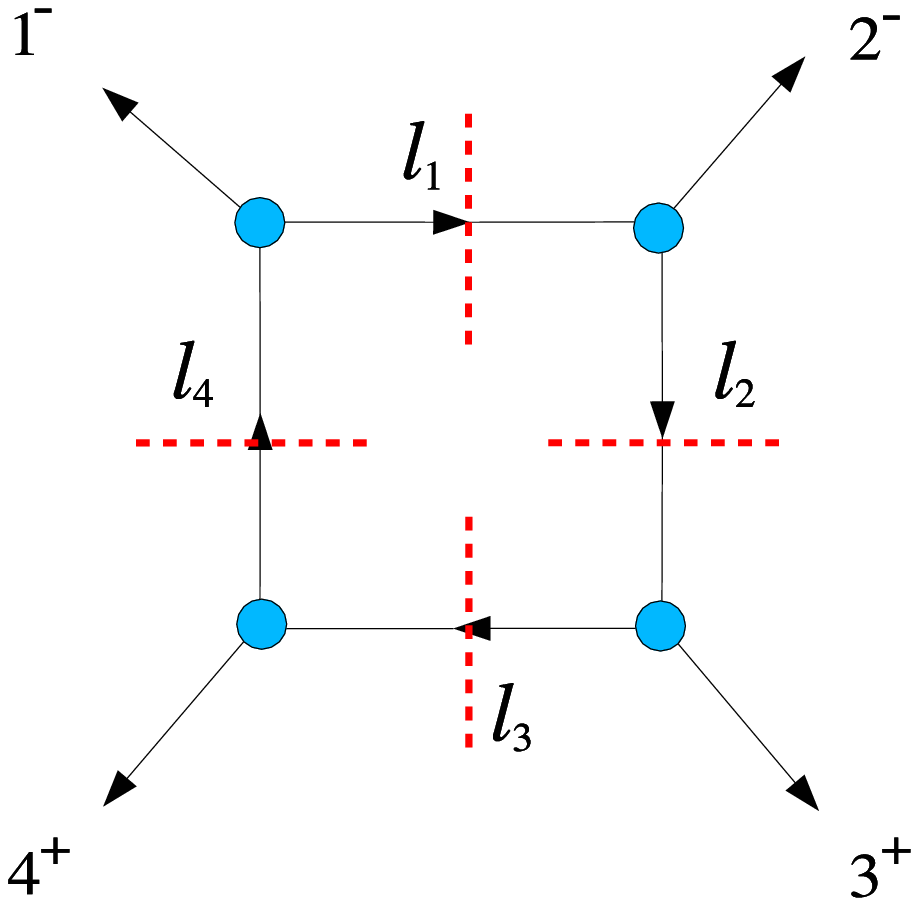}}
\end{center}
\caption{\it
The quadruple cut
for the amplitude $1^- 2^- 3^+ 4^+$.
}
\end{figure}
To begin with, we consider the quadruple cut
of the amplitude,  represented in Figure 5.
It is given by
\beq
\label{magilla}
\frac{\lan 1 \vert l_1 \vert q_1]}{[1 \, q_1]}\
\frac{\lan 2 \vert l_1 \vert q_2]}{[2 \, q_2]}\
\frac{\lan q_3 \vert l_3 \vert 3]}{\lan q_3 \, 3\ran}\
\frac{\lan q_4 \vert l_4 \vert 4]}{ \lan q_4 \, 4\ran  }
\ .
\eeq
By choosing $q_1=2$,  $q_2=1$,  $q_3=4$,  $q_4=3$,
\eqref{magilla} can be rewritten as
\beq
\label{czerny}
i \, {t \over s}\,  \cA_4^{\rm tree} \, \mu^4
\ ,
\eeq
where
\beq
\cA_4^{\rm tree} \ = \
i\frac{\lan 1 \, 2 \ran^3}{\lan 2 \, 3 \ran \lan 3 \, 4 \ran
\lan 4 \, 1 \ran}\ .
\eeq
Reinstating the four cut propagators and
integrating over the loop momentum, \eqref{czerny} gives
\beq
-\frac{\cA_4^{\rm tree} }{(4\pi)^{2-\epsilon}}
\bigg( \frac{t}{s}K_4
\bigg)
\ ,
\eeq
where $K_4$ is defined in \eqref{K4}.
\begin{figure}[ht]
\label{Figure4}
\begin{center}
\scalebox{0.70}{\includegraphics{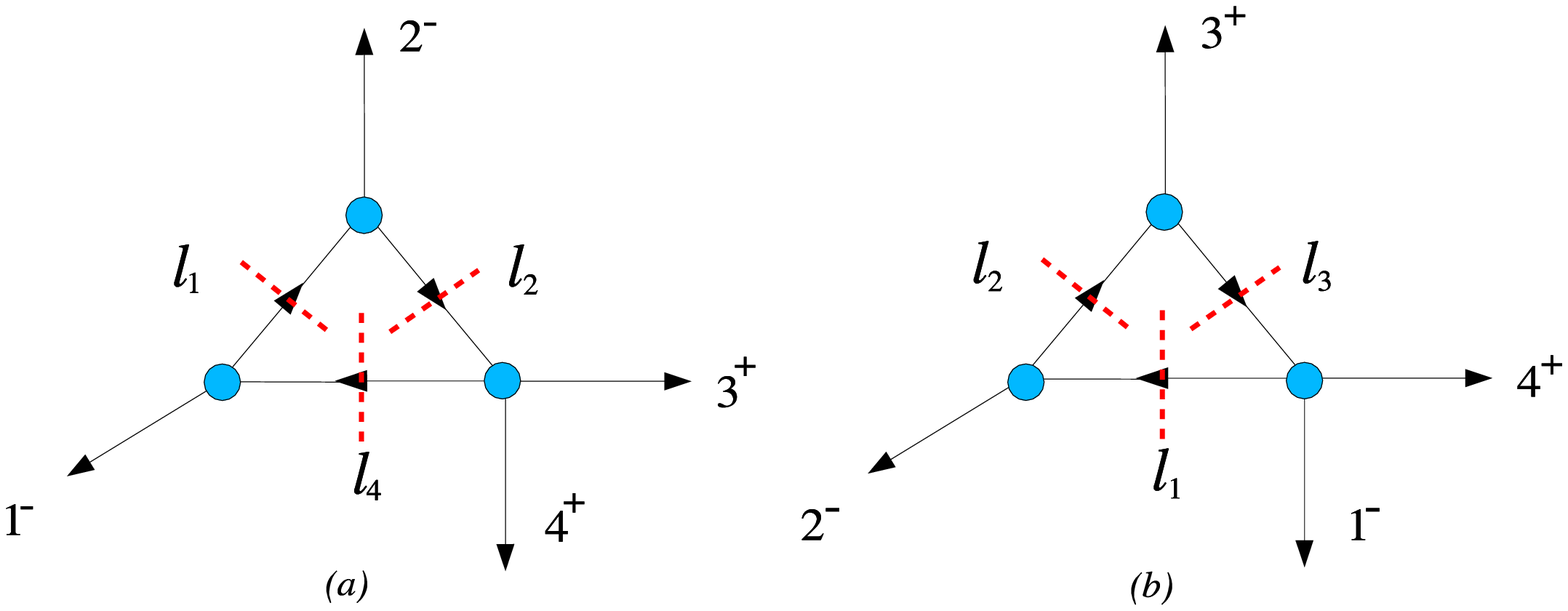}}
\end{center}
\caption{\it
The two inequivalent triple cuts
for the amplitude  $1^- 2^- 3^+ 4^+$.
}
\end{figure}

Next we consider triple cuts. We begin our analysis with
the triple cut in Figure 6{\it a}.
This yields
\beq
\frac{\mu^2 [3 \, 4]}{\lan 3 \, 4 \ran 2(l_2\cdot 3)}\
\frac{\lan 1 \vert l_1 \vert q_1]}{[1 \, q_1]}\
\frac{\lan 2 \vert l_1 \vert q_2]}{[2 \, q_2]}\
= \
-\mu^4 \frac{\lan 1\, 2 \ran [3 \, 4]}{[1 \, 2] \lan 3 \, 4 \ran}\
\frac{1}{2(l_2\cdot 3)}
\ ,
\eeq
which, upon reinstating the cut propagators
and performing the loop momentum integration
gives
\beq
-\frac{\cA_4^{\rm tree} }{(4\pi)^{2-\epsilon}}
\bigg( \frac{t}{s}K_4
\bigg)
\ .
\eeq
This function had already been detected with the
quadruple cut, as discussed earlier.

Next we move on to
consider the triple cut in Figure 6{\it b}.
This yields
\beq\label{adjacentmhv}
\frac{\lan 1 \vert l_3 \vert 4]^2}{2\, t \, (l_3 \cdot 4)}\
\frac{\lan 2 \vert l_1 \vert q_1]}{[2 \, q_1]}\
\frac{\lan q_2 \vert l_2 \vert 3]}{\lan q_2 \, 3 \ran}
\ .
\eeq
We can re-cast \eqref{adjacentmhv} as follows.
Firstly,  we write
\beq
\frac{\lan 1 \vert l_3 \vert 4 ]
\lan q_2 \vert l_3 \vert 3]}
{\lan q_2 \, 3 \ran} \ = \
\mu^2 \frac{\lan 1 \vert 4 \vert 3 ]} {\lan 3 \, 4 \ran} -
\frac{2 (l_3 \cdot 4) \lan 1 \vert l_3 \vert 3 ]} {\lan 3 \, 4 \ran}
\ ,
\eeq
and secondly
\beq
\frac{\lan 1 \vert l_3 \vert 4 ] \lan 2 \vert l_1
\vert q_1 ]} {[2 \, q_1]} \ =  \
\mu^2 \frac{\lan 2 \vert 1 \vert 4 ]} {[1 \, 2 ]} -
\frac{2 (l_3 \cdot 4)
\lan 2 \vert 1 \vert 4 ]}{[1 \, 2]} +
\frac{2(l_3 \cdot  4)
\lan 2 \vert l_3 \vert 4]}{[1 \, 2]}
\ .
\eeq
The expression (\ref{adjacentmhv}) becomes a sum of six terms
$T_i$, $i=1, \ldots , 6$, where
\beqa
\nonumber
T_1 & = & \frac{\lan 1\vert 4\vert 3]
\lan 2\vert 1\vert 4] \mu^4}{t \lan 3\, 4\ran [1\, 2] 2(l_3\cdot 4)}
\ ,
\nonumber \\[8pt]
T_2 & = & -\frac{\lan 1\vert 4\vert 3]
\lan 2\vert 1\vert 4] \mu^2}{t \lan 3\, 4\ran [1\, 2]}
\ ,
\nonumber \\[8pt]
T_3 & = &
\frac{\lan 1\vert 4\vert 3]
\lan 2\vert l_3\vert 4] \mu^2}{t \lan 3\, 4\ran [1\, 2]}
\ ,
\nonumber
\eeqa
\beqa
\nonumber
T_4 & = & -\frac{\lan 2\vert 1\vert 4]
\lan 1\vert l_3\vert 3]\mu^2}{t \lan 3\, 4\ran [1 \, 2]}
\ ,
\nonumber \\ [8pt]
T_5 & = &
\frac{\lan 2\vert 1\vert 4] \lan 1\vert l_3\vert 3]
2(l_3\cdot 4)}{t \lan 3 \, 4 \ran [1 \, 2]}
\ ,
\nonumber \\[8pt]
T_6 & = & -\frac{\lan 1 \vert l_3 \vert 3]
\lan 2 \vert l_3 \vert 4] 2(l_3\cdot 4)}{t \lan 3 \, 4 \ran [1 \, 2]}
\ .
\eeqa
Next we replace
the delta functions with propagators,
and integrate over the loop momentum.
To evaluate the integrals,
we use the linear, quadratic and cubic triangle integrals
in $4-2\eps$ dimensions listed in the Appendix.
The integration of the expressions gives
\beqa
T_1 & \to & -{\cA_4^{\rm tree} \over ( 4 \pi)^{2-\e}}
\bigg(\frac{t}{s}K_4 \bigg)
\ ,
\nonumber \\[8pt]
T_2 & \to & -{\cA_4^{\rm tree} \over ( 4 \pi)^{2-\e}}
\bigg(-\frac{t}{s}J_3(t)\bigg)
\ ,
\nonumber \\[8pt]
T_3 & \to &
-{\cA_4^{\rm tree} \over ( 4 \pi)^{2-\e}}
\bigg(\frac{t}{s}J_3(t)-\frac{1}{s}J_2(t)\bigg)
\ ,
\nonumber \\[8pt]
T_4 & \to &
-{\cA_4^{\rm tree} \over ( 4 \pi)^{2-\e}}
\bigg(-\frac{1}{s}J_2(t)\bigg)
\ ,
\nonumber \\ [8pt]
T_5 & \to &
-{\cA_4^{\rm tree} \over ( 4 \pi)^{2-\e}}
\bigg(\frac{t}{2s}I_2(t)+\frac{u}{s}I_3^{6-2\eps}(t)\bigg)
\ ,
\nonumber \\[8pt]
T_6 & \to &
-{\cA_4^{\rm tree} \over ( 4 \pi)^{2-\e}}
\bigg(-\frac{t}{4s}I_2(t)-
\bigg(\frac{3}{2s}+\frac{1}{t}\bigg)
I_2^{6-2\eps}(t)-\frac{u}{s}I_3^{6-2\eps}(t)\bigg)
\ .
\eeqa
We now use (A.26) in \cite{Bern:1995db}
relating $J_2(t)$ to $I_2(t)$
and $I_2^{6-2\eps}(t)$, and get
\beq
T_5+T_6 \to -\cA_4^{\rm tree}\bigg(\frac{1}{s}J_2(t)-
\frac{1}{t}I_2^{6-2\eps}(t)\bigg)
\ .
\eeq
Adding up the six $T_i$ terms, 
and including the usual factor of  two, 
we obtain
\beq\label{4ptresult--++}
-{2\cA_4^{\rm tree} \over ( 4 \pi)^{2-\e}}
\bigg( \frac{t}{s}K_4 -
\frac{1}{s}J_2(t) - \frac{1}{t}I_2^{6-2\epsilon}(t)
\bigg)
\ ,
\eeq
which precisely agrees with
\eqref{4pt--++}.

\subsection{The one-loop $-$$+$$-$$+$ amplitude}

Now we consider the one-loop amplitude with a complex scalar in the loop, $\cA^{\rm scalar}_4(1^-,2^+,3^-,4^+)$,
which is given by  \cite{Bern:1995db}%

\beqa\label{alternating}
\cA_4^{\rm scalar}(1^-,2^+,3^-,4^+) &=&
-2\frac{1}{(4\pi)^{2-\epsilon}}
\cA_4^{\rm tree}
\bigg(\frac{st}{u^2}K_4 - \frac{s^2t^2}{u^3}I_4^{6-2\eps}
+ \frac{st}{u^2}I_3^{6-2\eps}(t) 
\\ [8pt]\nonumber
 &+& \frac{st}{u^2}I_3^{6-2\eps}(s)  - \frac{st(s-t)}{u^3}J_3(t)
 - \frac{st(t-s)}{u^3}J_3(s)
+ \frac{s}{u^2}J_2(t) + \frac{t}{u^2}J_2(s)
\\ [8pt]\nonumber
 &+& \frac{s}{tu}I_2^{6-2\eps}(t) + \frac{t}{su}I_2^{6-2\eps}(s)
+\frac{ts^2}{u^3}I_2(t) + \frac{st^2}{u^3}I_2(s)
\bigg)
\ .
\eeqa

\begin{figure}[ht]
\label{figure-quad5}
\begin{center}
\scalebox{0.65}{\includegraphics{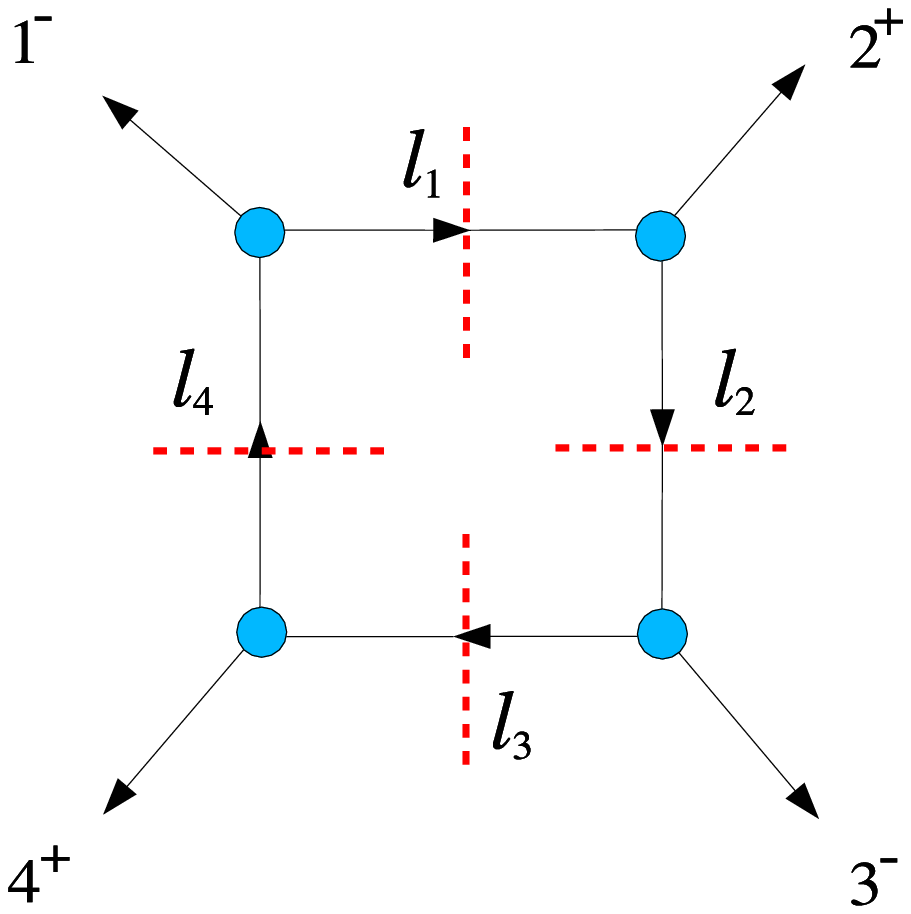}}
\end{center}
\caption{\it
The quadruple cut
for the amplitude $1^- 2^+ 3^- 4^+$.
}
\end{figure}

The relevant quadruple cut is represented in Figure 7, and
gives:
\beqa  \label{brumbrum2}
&&\frac{\lan 1\vert l_1\vert q_1]}{[1\, q_1]}\
\frac{\lan q_2\vert l_2\vert 2]}{\lan q_2\, 2\ran}\
\frac{\lan 3\vert l_3\vert q_3]}{[3 \, q_3]}\
\frac{\lan q_4\vert l_4\vert q_4]}{\lan q_4\, 4 \ran}\
\nonumber \\[8pt]
&&=\frac{1}{[1 \, 3] \lan 2\, 4\ran}\
\bigg(\lan 1\, 3\ran \mu^2 +\lan 1\, 2\ran
\lan 3\vert l_1\vert 2]\bigg)
\bigg([2 \, 4]\mu^2 -[3 \, 4] \lan 3\vert l_1\vert 2]\bigg)
\nonumber \\[8pt]
&&=i\cA_4^{\rm{tree}}\bigg(\frac{st\mu^4}{u^2}+
\frac{2s^2t\lan \vert l_1\vert 2]\mu^2}{
u^2 \lan 3\vert 1\vert 2]}+\frac{
s^3t\lan 3\vert l_1\vert 2]^2}{u^2\lan 3\vert 1\vert 2]^2}\bigg)
\ ,
\eeqa
where
\beq
\cA_4^{\rm tree} \ = \
i\frac{\lan 13\ran^4}{\lan 12\ran\lan23\ran
\lan 34\ran\lan 41\ran}
\ .
\eeq
Averaging over the two solutions of the quadruple cut we obtain
the following expression:
\beq
\label{boooh}
i\cA_4^{\rm{tree}}\bigg(\frac{st}{u^2}\
\mu^4+\frac{2s^2t^2}{u^3}\ \mu^2+\frac{s^3t^3}{2u^4}\bigg)
\ .
\eeq
After reinstating the four cut propagators and
integrating over the loop momentum, \eqref{boooh}
gives
\beq
\frac{1}{(4\pi)^{2-\eps}} \cA_4^{\rm{tree}}\bigg(
-\frac{st}{u^2} K_4-\frac{2s^2t^2}{u^3}
J_4-\frac{s^3t^3}{2u^4} I_4 \bigg)
\ .
\eeq
We now use the identity (A.26)
in \cite{Bern:1995db} ignoring functions
that do not have a quadruple cut to write this as
\beq
\frac{1}{(4\pi)^{2-\eps}} \cA_4^{\rm{tree}}\bigg(
-\frac{st}{u^2} K_4+\frac{s^2t^2}{u^3} I_4^{6-2\eps}\bigg)
\ .
\eeq

\begin{figure}[ht]
\label{triangle6}
\begin{center}
\scalebox{0.70}{\includegraphics{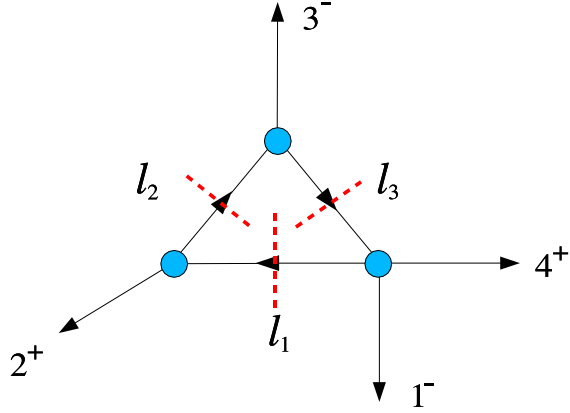}}
\end{center}
\caption{\it
The only independent  triple cut
for the amplitude  $1^- 2^+ 3^- 4^+$ (the others are obtained
from this one by cyclic relabeling of 
the external gluons).
}
\end{figure}

We now consider triple cuts.
There is only one independent triple cut, and
we consider, for instance,
the  triple cut in Figure 8,
which gives
\beq\label{nonadjacentmhv}
\frac{\lan 1 \vert l_3 \vert 4 ]^2} {2\, t \, (l_3\cdot 4)}\
\frac{\lan 3 \vert l_3 \vert q_2 ]} {[3 \, q_2]}\
\frac{\lan q_1 \vert l_1 \vert 2 ]} {\lan q_1 \, 2 \ran}
\ .
\eeq
Using straightforward spinor manipulations,
and taking into account properties of the cut momenta,
one finds that the above expression
may be expanded as a product of two sets of terms.
The first is
\beq
\frac{\lan 1 \vert l_3 \vert 4 ]
\lan 3 \vert l_3 \vert q_2 ]}{[3 \, q_2]}\ = \
\frac{\mu^2 \lan 3 \vert 1 \vert 4]}{[1 \, 3]}-
\frac{t\lan 3 \vert l_3 \vert 4]}{[1 \, 3]}+
\frac{2(l_3\cdot 4) \lan 3 \vert l_3 \vert 4]}{[1 \, 3]}
\ ,
\eeq
whereas the second is
\beq
\frac{\lan 1 \vert l_3 \vert 4 ]
\lan q_1 \vert l_1 \vert 2]}{\lan q_1 \, 2 \ran}
\ = \
\frac{\mu^2 \lan 1 \vert 4 \vert 2]}{\lan 2 \, 4 \ran}
+\frac{\lan 4 \vert 1 \vert 2] \lan 1 \vert l_3 \vert 4]}{
\lan 2 \, 4 \ran}-
\frac{2(l_3.4)\lan 1 \vert l_3 \vert 2]}{\lan 2 \, 4 \ran}
\ .
\eeq
The expression (\ref{nonadjacentmhv}) becomes then
a sum of nine terms $R_i$, $i = 1, \ldots , 9$, where
\beqa
\nonumber
R_1&=& \frac{\lan 1 \vert 4\vert 2] \lan 3\vert 1\vert 4] \mu^{4}}{
t [1 \, 3] \lan 2 \, 4\ran 2(l_3\cdot 4)}
\ ,
\nonumber \\[8pt]
R_2&=& \frac{\lan 4\vert 1\vert 2]
\lan 3\vert 1\vert 4] \lan 1\vert l_3\vert 4] \mu^{2}}{
t [1 \, 3] \lan 2 \, 4\ran 2(l_3\cdot 4)}
\ ,
\nonumber \\[8pt]
R_3&=& -\frac{\lan 3\vert 1\vert 4]
\lan 1\vert l_3\vert 2] \mu^2}{t [1 \, 3] \lan 2 \, 4\ran}
\ ,
\nonumber
\\[8pt]
R_4&=& -\frac{\lan 1\vert 4\vert 2]
\lan 3\vert l_3\vert 4] \mu^2}{[1 \, 3] \lan 2 \, 4\ran 2(l_3\cdot 4)}
\ ,
\nonumber \\[8pt]
R_5&=& -\frac{\lan 4 \vert 1 \vert 2]
\lan 3 \vert l_3 \vert 4]
\lan 1 \vert l_3 \vert 4]}{[1 \, 3] \lan 2 \, 4 \ran 2(l_3\cdot 4)}
\ ,
\nonumber \\[8pt]
R_6&=& \frac{\lan 3\vert l_3\vert 4]
\lan 1\vert l_3\vert 2]}{[1 \, 3] \lan 2 \, 4\ran}
\ ,
\nonumber \\[8pt]
R_7&=& \frac{\lan 1\vert 4\vert 2]
\lan 3\vert l_3\vert 4] \mu^2}{t [1 \, 3] \lan 2 \, 4\ran}
\ ,
\nonumber \\[8pt]
R_8&=& \frac{\lan 4\vert 1\vert 2]
\lan 3\vert l_3\vert 4]
\lan 1\vert l_3\vert 4]}{t [1 \, 3] \lan 2 \, 4\ran}
\ ,
\nonumber \\[8pt]
R_9&=& -\frac{\lan 3\vert l_3\vert 4]
\lan 1\vert l_3\vert 2] 2(l_3\cdot 4)}{t [1 \, 3] \lan 2 \, 4\ran}
\ .
\eeqa
The term $R_5$ becomes a quadratic box integral
when the three delta functions are replaced with propagators.
We can use the properties of the cut momenta
to re-write $R_5$ as a sum of 
terms which will give 
a box integral,
a linear box integral and a linear triangle integral
as follows,
\beq
R_5 =
-\frac{\lan 4\vert 1\vert 2] [4\vert 3 \, 1\vert 4] \mu^2}{
[1 \, 3]^2 \lan 2 \, 4\ran 2(l_3\cdot 4)}
+\frac{t \lan 4\vert 1\vert 2] [4\vert 3 \, l_3 \vert 4]}{
[1 \, 3]^2 \lan 2 \, 4\ran 2(l_3\cdot 4)}
-\frac{\lan 4\vert 1\vert 2] [4\vert 3 \, l_3 \vert 4]}{
[1 \, 3]^2 \lan 2 \, 4\ran}
\ .
\eeq
We now replace the delta functions
with propagators and integrate over the cut momenta.
Note that one must drop any terms
without cuts in the $t$-channel.
This must be used for all the
linear box integrals that appear above.
Using the results for the linear box and the
linear, quadratic and cubic triangle integrals
in $4-2\eps$ dimensions listed in the Appendix gives
\beqa
R_1 & \to &
-{\cA_4^{\rm tree} \over (4 \pi)^{2 - \e}}
\bigg(\frac{st}{u^2}K_4\bigg)
\ ,
\nonumber \\[8pt]
R_2 & \to &
-{\cA_4^{\rm tree} \over (4 \pi)^{2 - \e}}
\bigg(\frac{s^2t^2}{2u^3}J_4-\frac{s^2 t}{u^3}J_3(t)\bigg)
\ ,
\nonumber \\[8pt]
R_3 & \to &
-{\cA_4^{\rm tree} \over (4 \pi)^{2 - \e}}
\bigg(-\frac{st}{u^2}J_3(t)+\frac{s}{u^2}J_2(t)\bigg)
\ ,
\nonumber \\[8pt]
R_4 & \to &
-{\cA_4^{\rm tree} \over (4 \pi)^{2 - \e}}
\bigg(\frac{s^2t^2}{2u^3}J_4+\frac{st^2}{u^3}J_3(t) \bigg)
\ ,
\nonumber
\\[8pt]
R_5 & \to &
-{\cA_4^{\rm tree} \over (4 \pi)^{2 - \e}}
\bigg(\frac{s^2t^2}{u^3}J_4+\frac{s^3t^3}{2u^4}I_4+
\frac{s^2t^3}{u^4}I_3(t)+\frac{s^2t}{u^3}I_2(t)\bigg)
\ ,
\nonumber \\[8pt]
R_6 & \to & -{\cA_4^{\rm tree} \over (4 \pi)^{2 - \e}}
\bigg(\frac{st}{2u^2}I_2(t)\bigg)
\ ,
\nonumber \\[8pt]
R_7 & \to &
-{\cA_4^{\rm tree} \over (4 \pi)^{2 - \e}}
\bigg(\frac{s}{u^2}J_2(t) \bigg)
\ ,
\nonumber \\[8pt]
R_8 & \to & -{\cA_4^{\rm tree} \over (4 \pi)^{2 - \e}}
\bigg(-\frac{s^2}{u^2}I_3^{6-2\eps}(t) \bigg)
\ ,
\nonumber \\[8pt]
R_9 & \to &
 -{\cA_4^{\rm tree} \over (4 \pi)^{2 - \e}}
\bigg(-\frac{st}{4u^2}I_2(t)+\bigg(\frac{s}{2u^2}-
\frac{s^2}{u^2t}\bigg)I_2^{6-2\eps}(t)+
\frac{s^2}{u^2}I_3^{6-2\eps}(t)\bigg)
\ .
\eeqa
Now using (A.26) in \cite{Bern:1995db}, and
ignoring all terms without cuts in
the $t$-channel,  it is easy
to show that the sum of these nine terms leads to the result
\beqa\label{nonadjtcut}
\cA^{\rm t-cut}(1^-,2^+,3^-,4^+) &=&
-\frac{1}{(4\pi)^{2-\epsilon}}
\cA_4^{\rm tree}
\bigg(\frac{st}{u^2}K_4 - \frac{s^2t^2}{u^3}I_4^{6-2\eps}
+ \frac{st}{u^2}I_3^{6-2\eps}(t)
\\ [8pt]\nonumber
&-& \frac{st(s-t)}{u^3}J_3(t)
+ \frac{s}{u^2}J_2(t)
+ \frac{s}{tu}I_2^{6-2\eps}(t)
+ \frac{ts^2}{u^3}I_2(t)
\bigg)
\ .
\eeqa
Next, one must also include the corresponding terms
coming from the $s$-channel version
of the of triple cut in Figure 8.
This just yields \eqref{nonadjtcut}
with $t$ replaced by $s$.
Combining these two expressions,
without double-counting the box contributions
(which appear in both cuts), 
and including the usual factor of two, 
one precisely reproduces the amplitude
for this process \eqref{alternating}


\section{The $+$$+$$+$$+$$+$ amplitude}


The five-gluon all-plus one loop amplitude,
with a scalar in the loop,
is given by \cite{bdk9302280}
\beq\label{+++++}
\cA_5(1^+, 2^+, 3^+, 4^+, 5^+)  \ = \
{i\over 96 \pi^2 C_5}\bigg[
s_{12}s_{23} + s_{23}s_{34} + s_{34}s_{45} +
s_{45}s_{51} + s_{51}s_{12} + 4i
\eps(1234)\bigg] \ ,
\eeq
where
$C_5:=\lan 12 \ran \lan 23 \ran
\lan 34 \ran \lan 45 \ran \lan 51 \ran$ and
$ \e (abcd) := \e_{\m \n \r \s}\, a^\m b^\n c^\r d^\s$.

An expression for the five-gluon amplitude valid
to all orders in $\e$ appears in \cite{Bern:1996ja},
\beqa
\nonumber
&&
\cA_{5;1}^{\rm scalar} (1^+,2^+,3^+,4^+,5^+) \ = \
{ i \over   C_5}
{\eps(1-\eps) \over (4\pi)^{2-\eps} }
\biggl[
s_{23}s_{34} I_4^{(1),8-2\eps}
\, +\, s_{34}s_{45} I_4^{(2),8-2\eps}
\\ \nonumber \cr
&&
\qquad \qquad 
+\, s_{45}s_{51} I_4^{(3),8-2\eps}
 \, +\, s_{51}s_{12} I_4^{(4),8-2\eps}
\, +\, s_{12}s_{23} I_4^{(5),8-2\eps}
\\  \cr
&&
\qquad \qquad 
+ \, 4i(4-2\eps) \, { \eps(1234) }I_5^{10-2\eps}
\biggr] 
\ .
\label{+++++eps}
\eeqa
The result \eqref{+++++} is obtained from
\eqref{+++++eps} by taking the $\e \to 0 $ limit, where
\cite{Bern:1996ja}
\beq
\e (1-\e ) I_4^{8-2\e }
\rightarrow {1\over 6} \ ,
\qquad
\e(1-\e) I_5^{10-2\e} \rightarrow {1\over 24} \ ,
\qquad
\e(1-\e) I_6^{10-2\e} \rightarrow  0
\ .
\eeq
\begin{figure}[ht]
\label{figure-quad2}
\begin{center}
\scalebox{0.65}{\includegraphics{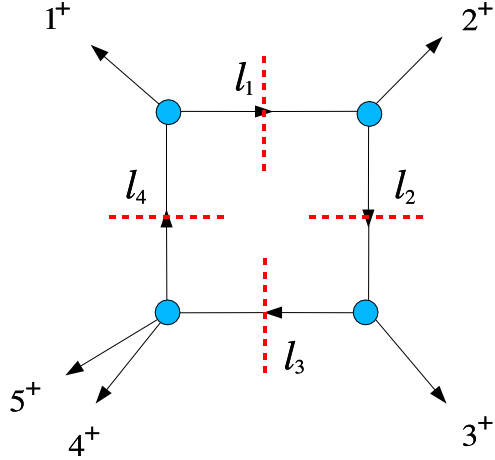}}
\end{center}
\caption{\it
One of the quadruple cuts
for the amplitude $1^+ 2^+ 3^+ 4^+5^+$.
}
\end{figure}

Here we will find that we can reproduce the full amplitude
using  only quadruple cuts in $4 - 2 \e$ dimensions.

Let us start by considering
the diagram in Figure 9, which represents the quadruple cut
where  gluons $4$ and $5$ enter the same tree amplitude.
The momentum constraints on this quadruple cut
are given by
\begin{eqnarray}\label{momcon2}\nonumber
&&l_1^2\, =\, l_2^2\, = \, l_3^2\, = \, l_4^2\, = \, \mu^2 \ ,
\\ [6pt]
&&l_1\, =\, l_4-k_1 \ , \
l_2\, =\, l_1-k_2\ , \
l_3\, = \, l_2-k_3\ , \
l_4\, =\, l_3-k_4-k_5
\ .
\end{eqnarray}
It will prove convenient to solve for the momentum $l_3$,
which we expand in the
basis of vectors
$k_1,k_2,k_3$ and $K$, where
$K$ is defined in
\eqref{Kthingy}.
One finds that the solution of \eqref{momcon2}
is given by%
\footnote{We notice that, had we solved for $l_1$, the solution
would have taken the form \eqref{elvira}  with the same
coefficients $a$, $b$, $c$, $d$ of
\eqref{dongiovanni} - but with $u$ defined by
$u = - s - t - (k_4 + k_5)^2.$}
\beq
l_3 \  =\ ak_1\, + \, bk_2\, + \, ck_3\, + \, dK
\ ,
\eeq
with
\begin{eqnarray}
\label{momsol}
&&a  \ =\ \frac{t}{2u}\ , \ b\ =\ -\frac{1}{2}\ ,
\ c \ =\ -1-\frac{s}{2u}\ ,
\\ [6pt]\nonumber
&&d \ =\ \pm \sqrt{-{st+4\mu^2u \over s t u^2}}
\ ,
\end{eqnarray}
where the kinematical invariants
$s$, $t$, $u$ are again defined by
\eqref{stu}, but now $s + t + u = (k_4 + k_5)^2$.

Considering the diagram in Figure 9,
the product of tree-level amplitudes entering the
quadruple cut can be written as
\beq
\label{expression1}
\frac{\lan q_1\vert l_1\vert 1]}{\lan q_1 1\ran}\
\frac{\lan q_2 \vert l_2\vert 2]}{\lan q_2 2 \ran}\
\frac{\lan q_3 \vert l_3\vert 3]}{\lan q_3 3 \ran}\
\frac{\m^2 \, [45] }{\lan 45\ran \, [ (l_3-k_4 )^2-\mu^2]}\
\ .
\eeq
Using \eqref{simplific},
and choosing $q_3 \!=\!2$, \eqref{expression1} can be recast as
\beqa\nonumber
& - &  \m^4 \,  {[12] \over \lan 12 \ran }
{[45] \over \lan 45 \ran }
{1 \over \lan 23 \ran}  \, { \lan 2 | l_3 | 3 ] \over
(l_3-k_4 )^2-\mu^2}
 \ = \
{\m^4 \over \lan 12 \ran \lan 23 \ran \lan 34 \ran
\lan 45 \ran \lan 51 \ran} { {\rm Tr_{-}} ( 5123l_3 4) \over
(l_3-k_4 )^2-\mu^2}
\\
& = &
- {\m^4 \over \lan 12 \ran \lan 23 \ran \lan 34 \ran
\lan 45 \ran \lan 51 \ran}
{ {\rm Tr_{+}} ( 123l_3 43)\, + \,  {\rm Tr_{+}} ( 123l_3 42)
\over
(l_3-k_4 )^2-\mu^2}
\ .
\label{expression2}
\eeqa
Using momentum conservation, and
\beq
{\rm Tr_{+}} ( abcd) \ = \ 2 \Big[ (ab) (cd) \, -  \, (ac) (bd)
\, +  \, (ad) (bc)\,  + \,  i \e (abcd) \Big]
\ ,
\eeq
it is easy to see that
\beq
{
{\rm Tr_{+}} ( 123l_3 43)\, + \,  {\rm Tr_{+}} ( 123l_3 42)
\over
(l_3 \cdot k_4) } \ = \ 4 (12) (23) \, - \,
4i { (34) \, \e( 12l_3 3)  \ - \ (12) \, \e (234l_3)
\over (l_3 \cdot k_4)}
\ .
\eeq
We set
\beq\label{expression4}
V(l_3)\  = \ i\eps(12l_33)(3 \cdot 4) \, - \,
i \eps(234l_3)(1 \cdot 2)
\ .
\eeq
Now we wish to sum the expression
\eqref{expression2}
over the solutions
\eqref{momsol}, including a factor of $1/2$.
Writing these solutions as
$l_3^\pm = x\pm y$, where $y$ contains the
term involving the momentum $K$,
it is straightforward to show that
\beq\label{expression5}
\frac{1}{2}\sum_{l_3^\pm}
{
{\rm Tr_{+}} ( 123l_3 43)\, + \,  {\rm Tr_{+}} ( 123l_3 42)
\over
(l_3 \cdot k_4) } \ = \
4\, (1\cdot 2)(2\cdot 3) \, - \,
4 {V (x) (x \cdot 4) \, - \, V (y) (y \cdot 4)
\over (x \cdot 4)^2 - (y \cdot 4)^2 }
\ ,
\eeq
and
\beq
{V (x) (x \cdot 4) \, - \, V (y) (y \cdot 4)
\over (x \cdot 4)^2 - (y \cdot 4)^2 }
\ = \
-{i\over 2} \m^2 \e (1234) \,
\bigg[\frac{1}{(l_3^+ \cdot 4)} \, + \,
\frac{1}{(l_3^- \cdot 4)}\bigg]
\ .
\eeq
Summarising, we have found that
\beqa
\nonumber
&& \frac{1}{2}\sum_{l_3^\pm}
{
{\rm Tr_{+}} ( 123l_3 43)\, + \,  {\rm Tr_{+}} ( 123l_3 42)
\over
(l_3 \cdot k_4) }  \ = \
4\, (1\cdot 2)(2\cdot 3) \, + \,
2i\m^2 \e (1234) \,
\bigg[\frac{1}{(l_3^+ \cdot 4)} \, + \,
\frac{1}{(l_3^- \cdot 4)}\bigg]
\\ 
&& \qquad \qquad \qquad \quad
=\   
s_{12} \, s_{23} \, - \, 4i \m^2 \e (1234) \,
\bigg[\frac{1}{(l_3^+ -k_4)^2 - \m^2} \, + \,
\frac{1}{(l_3^- - k_4)^2 - \m^2}\bigg]
\ .
\label{senzanome}
\eeqa
\mbox{} From \eqref{expression2}, we see that the full amplitude
in the quadruple cut is obtained by multiplying \eqref{senzanome}
by $- \m^4 / C_5$.
Next,
we lift the cut integral to a full Feynman integral, 
and get 
\beqa
\nonumber
&& - \, 2 {\m^4 \over C_5} \bigg[ 
s_{12} \, s_{23} \, - \, 4i \m^2 \e (1234) \,
\bigg(\frac{1}{(l_3^+ -k_4)^2 - \m^2} \, + \,
\frac{1}{(l_3^- - k_4)^2 - \m^2}\bigg)
\biggr] 
\ \longrightarrow 
\\ \nonumber 
&& 
\ \ \ \ \qquad 
- {i \over C_5 ( 4 \pi)^{2 - \e}}\, 
\biggl[ 
I_4^{(5), 4 - 2 \e} [ \m^4] \, s_{12} \, s_{23}
\, + \, 8 i I_5^{4 - 2 \e} [ \m^6] \, \e (1234) 
\biggr] 
\ 
\\
&&
\qquad 
 \ = \ 
 {i \over C_5} {\e (1-\e)\over ( 4 \pi)^{2 - \e}}  \,  
\biggl[ 
s_{12} \, s_{23}\, 
I_4^{(5), 8-2 \e} \, + \, 4i \, (4 - 2 \e) \,  \e (1234)
I_5^{10-2\e}
\biggr]
\ , 
\label{leporello}
\eeqa
where the factor of $2$ in the first line of 
\eqref{leporello} comes 
from adding, as usual, the two possible quadruple cuts 
of the amplitude 
(which are equal, since they are obtained one 
from the other by simply flipping all the 
internal ``scalar helicities'').

Let us now discuss  the result we have found. 
The first term in the last line of 
\eqref{leporello}  gives 
the $s_{12} s_{23}$ term in  \eqref{+++++eps}.
The other quadruple cut diagrams, which 
come from cyclic relabelling of 
the external legs, will similarly generate
the other $\eps (1234)$-independent terms in \eqref{+++++eps}.
Finally, the $\e (1234)$ term in \eqref{leporello} -- 
a pentagon integral term --  matches 
the $\e (1234)$ term in \eqref{+++++eps}.

Thus we have shown that the five gluon amplitude
$+$$+$$+$$+$$+$
may be reconstructed directly using quadruple cuts
in $4 - 2 \e$ dimensions.

\newpage


                \section*{Acknowledgements}


This work was partially supported
by a Particle Physics and Astronomy
Research Council award {\it M Theory, string theory and duality},
and a European Union Framework 6 Marie Curie Research
Training Network
grant {\it Superstrings}.
AB would like to thank the
Albert-Einstein-Institute in Golm and the
School of Physics and Astronomy at Tel-Aviv University
for hospitality during various stages of this work.
GT would like to thank Marco Matone and the
Physics Department at the University of Padova
for hospitality during the initial stage of this work.
We would like to thank
James Bedford, Emil Bjerrum-Bohr, Stefano Catani,
Chong-Sun Chu, Dave Dunbar, Nigel Glover, Massimiliano Grazzini,
Harald Ita, Valya Khoze, David Kosower, Marco Matone and
Sanjaye Ramgoolam for stimulating conversations.


\startappendix


\Appendix{Tensor Integrals}


In this section we summarise the tensor bubble,
tensor triangle and
tensor box integrals used in this paper.

The scalar $n$-point integral functions
in $D=4+2m-2 \eps$ dimensions are defined as
\beqa
I^{D}_{n} \equiv I^{D}_{n}[1] & = &
i (-1)^{n+1}(4 \pi)^{D/2} \int \frac{d^{D}L}{(2 \pi)^{D}}
\frac{1}{L^{2} (L-p_{1})^{2}
\cdots (L-\sum_{i=1}^{n-1} p_{i})^{2}} \\
& = &
\frac{i (-1)^{n+1}}{\pi^{2+m-\eps}} \int
\frac{d^{4+2m}l \, d^{-2\eps}\mu}{(l^{2}-\mu^{2})
((l-p_{1})^{2}-\mu^{2})
\cdots ((l-\sum_{i=1}^{n-1} p_{i})^{2}-\mu^{2})}
\ .
\nonumber
\eeqa
The higher dimensional integral functions are related
to $4-2\eps$ dimensional integrals with a factor $\mu^{2m}$
inserted in the integrand. For $m=1,2$ one finds
\be
I_{n}[\mu^{2}]\equiv J_{n} \ = \
(-\eps) I_{n}^{6-2 \eps}  \,\, , \,\, \mathrm{and} \,\,
I_{n}[\mu^{4}]\equiv K_{n}  = \
(-\eps)(1 - \eps) I_{n}^{8-2 \eps} \, .
\ee

In our paper we encounter bubble functions with $m=0,1$,
triangles with one massive external line and $m=0,1$,
and boxes with four massless external lines and $m=0,1,2$:
\beqa
&& I_{2}(P^{2})\ =\
\frac{r_{\Gamma}}{\eps(1-2 \eps)} (-P^{2})^{-\eps}
\,\, , \qquad
I_{2}^{6-2 \eps}(P^{2})\ =\
-\frac{r_{\Gamma}}{2 \eps(1-2 \eps)(3-2\eps)}
(-P^{2})^{1-\eps}
\,\, , \nonumber \\
&& I_{3}(P^{2})\ =\
\frac{r_{\Gamma}}{\eps^{2}} (-P^{2})^{-1-\eps}
\qquad \,\, , \qquad
I_{3}^{6-2 \eps}(P^{2})\ =\
-\frac{r_{\Gamma}}{2 \eps(1- \eps)(1-2\eps)}
(-P^{2})^{-\eps}
\,\, , \nonumber \\
&&I_{4}\ =\
-\frac{r_{\Gamma}}{st} \left\{ -\frac{1}{\eps^{2}} \left[
(-s)^{-\eps}+(-t)^{-\eps}\right] +\frac{1}{2}
\log^{2}\left( \frac{s}{t} \right) +\frac{\pi^{2}}{2} \right\}
+ \mathcal{O}(\eps) \,\, ,  \nonumber \\
&&(-\eps)I_{4}^{6-2\eps}\ =\
0 + \mathcal{O}(\eps) \,\, ,
\qquad (-\eps)(1-\eps)I_{4}^{8-2\eps}\ =\
-\frac{1}{6} + \mathcal{O}(\eps)
\,\, .
\eeqa
Note that the expressions for the bubbles
and triangles are valid to all orders in $\eps$,
whereas for the box functions we have only kept
the leading terms which contribute up to
$\mathcal{O}(\eps^{0})$ in the amplitudes.

\begin{figure}[ht]
\label{Feynman}
\begin{center}
\scalebox{0.65}{\includegraphics{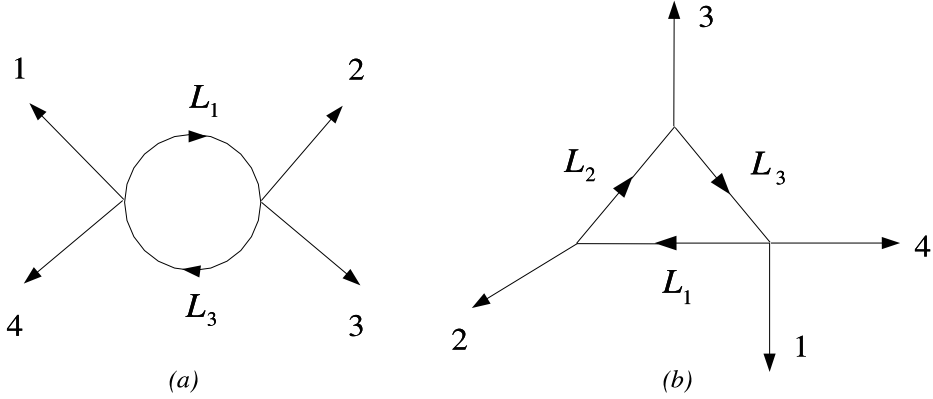}}
\end{center}
\caption{\it
Kinematics of the bubble and triangle integral functions
studied in this Appendix.
}
\end{figure}

We now move on to  present the result of the PV reduction for
various tensor integrals which are relevant for this paper.
Note that the expressions are presented
in terms of scalar $n$-point integral functions $I_{n}^\cD$
in various dimensions $\cD$, specifically in terms of
$I_n$, $I_n^{6-2\eps}$ and $I_n^{8-2\eps}$
in $4-2\eps$, $6-2\eps$ and  $8-2\eps$ dimensions, respectively.
The expressions are valid to all orders in $\eps$, if $I_{n}$,
$I_n^{6-2\eps}$ and $I_n^{8-2\eps}$
are evaluated to all orders, and
the PV reductions have been performed
in a fashion that naturally leads to coefficients
without explicit $\eps$ dependence
(the reader may consult \cite{bc9503236} for more details
on this particular variant of PV reductions).

For the linear and two-tensor bubbles  we have
(see Figure 10{\it a}):
\beqa
\label{AllPVs}
&&
I_2\big[ L_{3}^{\m} \big]
\ = \
-\frac{1}{2} I_2 (p_2+p_3)^\mu\ ,
\\ [8pt]
&&
I_2\big[ L_{3}^{\m} L_{3}^{\n}\big]
\ = \
  -\frac{1}{2} I_2^{6-2\eps}\delta^{\mu\nu}_{[4-2\eps]}
                   + \bigg(  \frac{1}{4} I_2 +
\frac{1}{2t} I_2^{6-2\eps}   \bigg) (p_2+p_3)^\mu
                   (p_2+p_3)^\nu\ .
%
\eeqa

For the linear, two- and three-tensor triangles
(see Figure 10{\it b}):
\beqa
&&
I_3\big[ L_{3}^{\m} \big]
\ = \
- \frac{1}{t} I_2 p_2^\mu +
                          \bigg( -I_3 +
 \frac{1}{t} I_2 \bigg) p_3^\mu\ ,
\\ [8pt]\nonumber
&&
I_3 \big[ L_{3}^{\m}L_{3}^{\n} \big]
\ = \
\frac{1}{2t} I_2 p_2^\mu p_2^\nu
                   + \bigg( \frac{1}{t} I_3^{6-2\eps} +
\frac{1}{2t} I_2 \bigg)
                   \bigg(p_2^\mu p_3^\nu+p_2^\nu p_3^\mu \bigg)
          \\ [8pt] && \qquad\qquad\qquad\qquad
           + \bigg( - \frac{3}{2t} I_2 +
I_3 \bigg) p_3^\mu p_3^\nu
                                      -\frac{1}{2} I_3^{6-2\eps} \delta^{\mu\nu}_{[4-2\eps]}\ ,
\\ [8pt]\nonumber
&&
I_3 \big[ L_{3}^{\m}L_{3}^{\n}L_{3}^{\rho} \big]
\ = \
                                   - \bigg( \frac{1}{4t} I_2 +
\frac{1}{2t^2} I_2^{6-2\eps}   \bigg)
\bigg(p_2^\mu p_2^\nu p_2^\rho \bigg)
\\ [8pt]\nonumber &&\qquad
-     \bigg(  \frac{1}{4t} I_2 +
\frac{3}{2t^2} I_2^{6-2\eps}  \bigg)
       \bigg( p_2^\mu p_2^\nu p_3^\rho +
p_2^\mu p_3^\nu p_2^\rho + p_3^\mu p_2^\nu p_2^\rho \bigg)
  \\ [8pt]\nonumber  && \qquad +
\bigg( -  \frac{1}{4t} I_2 + \frac{3}{2t^2} I_2^{6-2\eps}
- \frac{2}{t} I_3^{6-2\eps} \bigg)
\bigg(  p_2^\mu p_3^\nu p_3^\rho +
p_3^\mu p_3^\nu p_2^\rho + p_3^\mu p_2^\nu p_3^\rho \bigg)
 \\[8pt] \nonumber
&& \qquad +
\bigg(   \frac{7}{4t} I_2 + \frac{1}{2t^2} I_2^{6-2\eps}
   -  I_3 \bigg)  \bigg(  p_3^\mu p_3^\nu p_3^\rho  \bigg)
 +     \frac{1}{2t} I_2^{6-2\eps}
        \bigg(  \delta^{\mu\nu} p_2^\rho +
\delta^{\mu\rho} p_2^\nu+  \delta^{\rho\nu} p_2^\mu \bigg)
 \\ [8pt]
&& \qquad
+ \bigg(  - \frac{1}{2t} I_2^{6-2\eps}+
\frac{1}{2} I_3^{6-2\eps} \bigg)
 \bigg(  \delta^{\mu\nu} p_3^\rho +
\delta^{\mu\rho} p_3^\nu+ \delta^{\rho\nu} p_3^\mu \bigg)
\ .
\eeqa

Finally, for the linear box:
\beqa\label{linearbox}
I_4 \big[ L_{3}^{\m} \big]
&  = &
              \bigg( \frac{t}{2u} I_{4}\, - \, \frac{1}{u}
\big( I_{3}(t) - I_{3}(s)\big)\bigg) p_1^{\mu}
              -\frac{1}{2} I_{4} p_{2}^{\mu}
\nonumber  \\  [8pt] & + &
\bigg( \frac{t-u}{2u} I_{4}
\,  - \,
\frac{1}{u} \big( I_{3}(t) - I_{3}(s)\big)\bigg) p_3^{\mu} \ ,
\eeqa
where, as usual, $I_{n}^{D}$
denote $D$-dimensional scalar $n$-point integral functions,
$s:= (p_1 + p_2)^2$, $t:= (p_2 + p_3)^2$,
$ u := (p_1 + p_3)^2$,
and
$I_{n}$ is an abbreviation for the
$(4 \!-\!2 \eps)$-dimensional integral functions.
%


\end{document}